\documentclass[aps, pra, reprint, superscriptaddress]{revtex4-1}

\usepackage[utf8]{inputenc}
\usepackage[english]{babel}

\usepackage[T1]{fontenc}
\usepackage{lmodern}

\usepackage{amsmath}
\usepackage{amssymb}
\usepackage{dsfont}
\usepackage{graphicx}
\usepackage{url}
\usepackage{color}
\usepackage{soul}
\usepackage{afterpage}
\usepackage[verbose]{placeins}

\newcommand{\tr}{\mathrm{Tr}}
\newcommand{\im}{\mathrm{i}}
\newlength{\graphwidth}
\newcommand{\sys}{\mathrm{S}}
\newcommand{\bath}{\mathrm{B}}

\newcommand{\ddt}{\frac{\mathrm{d}}{\mathrm{d}t}}
\newcommand{\il}[3]{\int_{#1}^{#2}\mathrm{d}#3\,}
\newcommand{\hc}{\mathrm{hc.}}
\newcommand{\bra}[1]{\langle #1 |}
\newcommand{\ket}[1]{| #1 \rangle}
\newcommand{\proj}[1]{\ket{#1}\bra{#1}}
\newcommand{\scp}[2]{\bra{#1} #2 \rangle }
\newcommand{\derivD}{\mathrm{d}}
\newcommand{\ybf}{\mathbf{y}}
\newcommand{\zbf}{\mathbf{z}}
\newcommand{\kbf}{\mathbf{k}}

\begin{document}

\title{Exact open quantum system dynamics using the Hierarchy of Pure States (HOPS)}
\author{Richard Hartmann}
\email{richard.hartmann@tu-dresden.de}
\affiliation{Institut für Theoretische Physik, Technische Universität Dresden, D-01062 Dresden, Germany}
\author{Walter T. Strunz}
\email{walter.strunz@tu-dresden.de}
\affiliation{Institut für Theoretische Physik, Technische Universität Dresden, D-01062 Dresden, Germany}

\begin{abstract}
We show that the general and numerically exact \emph{Hierarchy of Pure States} method (HOPS) is very well applicable to calculate the reduced dynamics of an open quantum system. In particular we focus on environments with a sub-Ohmic spectral density (SD) resulting in an algebraic decay of the bath correlation function (BCF).
The universal applicability of HOPS, reaching from weak to strong coupling for zero and non-zero temperature, is demonstrated by solving the spin-boson model for which we find perfect agreement with other methods, each one suitable for a special regime of parameters.
The challenges arising in the strong coupling regime are not only reflected in the computational effort needed for the HOPS method to converge but also in the necessity for an importance sampling mechanism, accounted for by the non-linear variant of HOPS. In order to include non-zero temperature effects in the strong coupling regime we found that it is highly favorable for the HOPS method to use the zero temperature BCF and include temperature via a stochastic Hermitian contribution to the system Hamiltonian. 
\end{abstract}

\maketitle

\section{Introduction}

Including environmental effects when calculating the dynamics of quantum systems has been and still is a challenging task. As perfect isolation is an ideal concept, essentially any quantum system is exposed to environmental influences. 
To treat these influences, numerous approaches\cite{breuer_theory_2007, weiss_quantum_2008, rivas_open_2012, strunz_open_1999, suess_hierarchy_2014, tanimura_two-time_1989, tanimura_stochastic_2006, makri_tensor_1995, orth_nonperturbative_2013, meyer_multi-configurational_1990, beck_multiconfiguration_2000, wang_multilayer_2003} have been developed in many different contexts such as quantum optics, chemical and solid states physics, and also in cosmology. 
Non-perturbative approaches for general open systems that are used in the demanding parameter regime are for example: the quasi adiabatic path integral method (QUAPI\cite{makri_tensor_1995, thorwart_dynamics_2004, nalbach_ultraslow_2010}), variants of the time dependent Hartree method\cite{beck_multiconfiguration_2000} (e.g. ML-MCTDH\cite{wang_multilayer_2003}) and the hierarchical equations of motion (HEOM\cite{tanimura_two-time_1989, tanimura_stochastic_2006, tanimura_reduced_2014}). 
Despite the accuracy of these methods, computational limits occur in certain regimes. 
In case of the QUAPI method, the memory time of the environment must not be too long (sufficiently fast decay of the bath correlation function) in comparison to the resolution of the system dynamics. 
The wave function based ML-MCTDH method allows to efficiently treat very high dimensional quantum systems and therefore suits first principle calculations for open quantum system dynamics\cite{wang_coherent_2008, wang_coherent_2010} with a discretized environment. To deal with non-zero temperature initial conditions, the thermal average can be calculated using Monte Carlo sampling\cite{matzkies_accurate_1998, wang_quantum-mechanical_2006}.
Concerning HEOM, a representation of the bath correlation function (BCF) in terms of exponentials is required. 
Given a spectral density, the usual approach, where such an exponential form is generated via a Meier-Tannor (MT) decomposition\cite{meier_non-markovian_1999}, is very challenging to solve for low temperatures. Furthermore it has been shown that the MT decomposition poses difficulties when treating sub-Ohmic SD\cite{liu_reduced_2014, tang_extended_2015}. To bypass the MT decomposition more suitable representations of the BCF in terms of some special parametrization can be found, which allow to treat -- at least in principal -- any SD at any temperature. For example a HEOM based approach has been successfully used to study quantum impurity systems at low temperature\cite{li_hierarchical_2012, cheng_time-dependent_2015}. Further, an extended version of HEOM (eHEOM\cite{tang_extended_2015}) allows to treat sub-Ohmic environments at high temperatures \cite{tang_extended_2015} as well as zero temperature in the very demanding strong coupling regime\cite{duan_zero-temperature_2017}.

A stochastic state vector based alternative -- applicable to zero and non-zero temperature environments, also in the strong coupling regime -- is the hierarchy of pure states (HOPS) method, first introduced by Süß et al.\cite{suess_hierarchy_2014} and successfully used to calculate 2D electronic spectra\cite{zhang_non-perturbative_2016}. The method is based on the non Markovian quantum state diffusion (NMQSD) formulation for open quantum system dynamics\cite{diosi_non-markovian_1997, strunz_open_1999}. Here we introduce two new aspects for HOPS. First, we generate the exponential form of the BCF by a direct optimization procedure in the time domain of the BCF (similar to the recent work of Duan et al.\cite{duan_zero-temperature_2017}). In this way we can assure that for a given finite time interval the BCF approximated by exponentials mimics the decay of the exact BCF correctly. In particular we find highly accurate approximations for the zero temperature (sub-) Ohmic BCF. And second, thermal initial environmental states are modeled stochastically such that the truncation level of the hierarchy, and with that also the numeric effort, is temperature independent. By contrast, the truncation level is affected by the coupling strength which makes the strong coupling regime most challenging. The stochastic nature of the HOPS method can be coped with using straight forward numeric parallelization. 

The implementation of these two new aspects gives rise to consider HOPS as a generally applicable method in the sense that once an acceptable fit of the BCF at zero temperature was found, any thermal initial state can be dealt with.

To benchmark HOPS we solve the spin-boson model. In the weak coupling limit and for a fast decaying BCF the dynamics gained from HOPS is compared with calculations employing the quantum optical master equation and its correct extension to a sub-Ohmic environment (as explained in \cite{noh_<pre>consistent_2014}). Furthermore, when the spin is influenced mainly by classical noise induced by a high temperature environment, HOPS is compared with HEOM utilizing a MT decomposition and eHEOM. We confirm results obtained by the eHEOM method\cite{tang_extended_2015}, which circumvents the representation of the BCF in terms of exponentials, stating the inaccuracy of the MT decomposition for sub-Ohmic SDs. For special parameters we also note deviations between HOPS and eHEOM highlighting the approximative nature of simply using classical noise even at room temperature. In the strong coupling regime the comparison is drawn to the ML-MCTDH method at zero temperature. In this highly non trivial regime HOPS is also used to calculate spin dynamics for thermal initial environmental states.

\section{The HOPS method}

For reasons of readability the main ideas underlying the HOPS method (see \cite{suess_hierarchy_2014}) are presented here. Additionally, details concerning the approximation of the BCF, thermal initial conditions and the stochastic process sampling are elucidated.

\subsection{Sketch of the derivation}

\label{sec_HOPS_derivation}

Starting point is the usual system+bath Hamiltonian with (bath-)linear coupling between an arbitrary system $H_\sys$ and a bath consisting of a set of harmonic oscillators in the interaction picture with respect to the bath ($\hbar = 1$):
\begin{equation}
  H = H_\sys + \sum_\lambda L^\dagger g_\lambda e^{-\im \omega_\lambda t} a_\lambda + \sum_\lambda L g^\ast_\lambda e^{\im \omega_\lambda t} a^\dagger_\lambda
\end{equation}
Here, $L$ -- could, but not need to be self-adjoint -- denotes an arbitrary operator acting on the system Hilbert space and $a_\lambda$ ($a^\dagger_\lambda$) is the bosonic annihilation (creation) operator acting on the environmental mode with index $\lambda$. The reduced density matrix (RDM), which allows to calculate the expectation value for any observable on the system side, follows from the partial trace over the environmental degrees of freedom of the total state $\ket{\Psi}$. When expressing the trace in terms of Bargmann coherent states, which are unnormalized eigenstates of the annihilation operator defined as $\ket{z} = e^{za^\dagger}\ket{0}$, the integral, with its exponential weights, can be interpreted in a Monte Carlo sense
\begin{multline}
  \rho_\sys = \tr_\bath \proj{\Psi} = \int \mathrm{d}^2 \zbf e^{-|\zbf|^2}\scp{\zbf}{\Psi}\scp{\Psi}{\zbf} \\ = \mathrm{Mean}( \ket{\psi(\zbf^\ast)}\bra{\psi(\zbf)})
\end{multline}  
This allows to read the RDM as an average over pure stochastic states vectors $\psi(\zbf^\ast) := \scp{\zbf}{\Psi}$. In that way the coherent state labels $z_\lambda$ turn into complex valued Gaussian distributed random variables. Note, the bold faced $\zbf$ is shorthand notation for the vector $(z_1, z_2, ... z_\lambda, ...)$ and $\mathrm{d}^2 z_\lambda = \pi^{-1} \mathrm{d} \,\mathrm{Re}(z_\lambda) \;\mathrm{d}\, \mathrm{Im}(z_\lambda)$.

For an initial product state of the form $\ket{\Psi_0} = \ket{\psi}_\sys \ket{\mathbf{0}}_\bath$ (zero temperature bath) the time evolution of the stochastic state vector, following from the Schrödinger equation, reads
\begin{equation}
  \partial_t \psi_t[z^\ast] = \Big[-\im H_\sys + L z^\ast_t - L^\dagger \il{0}{t}{s} \alpha(t-s) \frac{\delta}{\delta z^\ast_s}\Big]\psi_t[z^\ast]
  \label{eqn:nmqsd}
\end{equation}
which is the non-Markovian quantum state diffusion equation\cite{strunz_open_1999} (NMQSD). The stochasticity of the coherent state labels $z_\lambda$ is contained in the scalar complex valued stochastic process $z^\ast_t = -\im \sum_\lambda g^\ast_\lambda z^\ast_\lambda e^{\im \omega_\lambda t}$ which turns the stochastic state vector into a functional of that stochastic process:  $\psi(\zbf^\ast) \rightarrow \psi_t[z^\ast]$. Crucially, the statistics of $z^\ast_t$ can also be expressed in terms of the zero temperature BCF $\alpha(t-s) := \sum_\lambda |g_\lambda|^2 e^{-\im \omega_\lambda (t-s)}$
\begin{equation}
  \langle z_t \rangle = \langle z_t z_s \rangle = 0 \qquad
  \langle z_t z^\ast_s \rangle = \alpha(t-s)
\end{equation}
implying that the knowledge of the BCF is sufficient to propagate the stochastic state vectors.

The convolution term including the functional derivative poses difficulties when proceeding without any approximation. However, the problematic term can be traded for a set of auxiliary states organized in a hierarchical structure. 

To proceed, a BCF written as a sum of exponentials
\begin{equation}
  \alpha(\tau) = \sum_{j=1}^N G_j e^{-W_j \tau} \qquad G_j, W_j \in \mathds{C}
  \label{eqn_bcf}
\end{equation}
is assumed, similar to other methods for density operators\cite{meier_non-markovian_1999, tanimura_stochastic_2006}. For a spectral density (SD) of Lorentzian shape this holds true exactly, whereas for other SD the validity of approximating the BCF in terms of a sum of exponentials has to be checked. 

Depending on the number of exponentials $N$ the convolution term in eq \ref{eqn:nmqsd} splits into $N$ terms. Each of these terms is viewed as an unknown vector named auxiliary state $\psi^\kbf$ and labeled with the $N$ dimensional index vector $\kbf$. For example, $(1,0,0...)$ is the index for the first auxiliary state, $(0,1,0 ...)$ for the second and so on. The evolution equation of the auxiliary states involves again the convolution term, resulting in auxiliary states for the auxiliary states. The so called auxiliary states of the second level have an index $\kbf$ with $\sum_{j=1}^N k_j = 2$. The index $\kbf = \mathbf{0}$ labels the stochastic state vectors  $\psi_t[z^\ast]$ which is the primary object of interest. It turns out that the influence of the high level auxiliary states on the zeroth level decreases as the level increases. This justifies the truncation of the hierarchy, allowing for the numeric integration of the following set of first order differential equations:
\begin{multline}
    \partial_t \psi_t^{\kbf} = \left[z^\ast_t L - \sum_{j=1}^N k_j W_j - \im H_\sys\right]\psi_t^{\kbf} \\
     + L \sum_{j=1}^N G_j k_j \psi_t^{\kbf - \mathbf{e}_j} - L^\dagger \sum_{j=1}^N \psi_t^{\kbf + \mathbf{e}_j}
     \label{eqn:hops} 
\end{multline}
Truncating the hierarchy at level $k_\mathrm{max}$ means that for each auxiliary state with level $k = \sum_j k_j = k_\mathrm{max}$ the dependence on higher level auxiliary states is simply neglected.
Solving HOPS refers to the task of solving the above set of differential equations for various realizations of the stochastic process $z^\ast_t$ and averaging over the projectors of $\psi^\mathbf{0}_t \equiv \psi_t[z^\ast]$.

It is important to point out that the actual implementation of HOPS extends eq \ref{eqn:hops} such that each sample has the same weight\cite{suess_hierarchy_2014}. Such an importance sampling follows from the nonlinear NMQSD Equation\cite{strunz_open_1999, de_vega_non-markovian_2005} where the environmental dynamics expressed in terms of the Husimi function takes the role of a time dependent sampling weight. So whenever the environment changes significantly, the nonlinear HOPS method is needed for efficient convergence of the RDM. Although the impact might be very significant, the actual modification to HOPS is minor, leaving the hierarchical structure and the dependence on the representation of the BCF in terms of a sum of exponentials unchanged.

\subsection{Exponential form of the bath correlation function}

In their standard form both HOPS and also HEOM\cite{tanimura_stochastic_2006} require a BCF that can be expressed as a sum of exponentials. For BCFs of different kind the required property may be approximately established by expressing the SD in terms of Lorentzians as proposed by Meier and Tannor\cite{meier_non-markovian_1999}. For practical purposes, this, however, requires high temperatures. Also this procedure should be used cautiously for BCFs with special asymptotic behavior like a diverging reorganization energy\cite{liu_reduced_2014}. Concerning HEOM, various extensions to reach for low temperatures and a broader class of SD have been proposed. For example the hybrid method sHEOM\cite{moix_hybrid_2013}, which combines a stochastic unraveling of the real part of the BCF with a deterministic treatment of the imaginary part by HEOM. In that way low temperatures are feasible. Alternatively, to cope with more general structured baths, the eHEOM method \cite{tang_extended_2015, duan_zero-temperature_2017} expands the BCF in terms of an arbitrary complete set of functions which results in additional couplings between the hierarchy branches.

In this work we follow a different path. From the NMQSD Equation (eq \ref{eqn:nmqsd}) it can be deduced that the time evolution of the stochastic state vector $\psi_t[z^\ast]$ only depends on the BCF over the interval $[0, t]$. Therefore, to evaluate the reduced dynamics up to time $t$ it is sufficient to have an expression of the BCF in terms of exponentials for that particular time interval only. 

\begin{figure*}
  \includegraphics[width=\graphwidth]{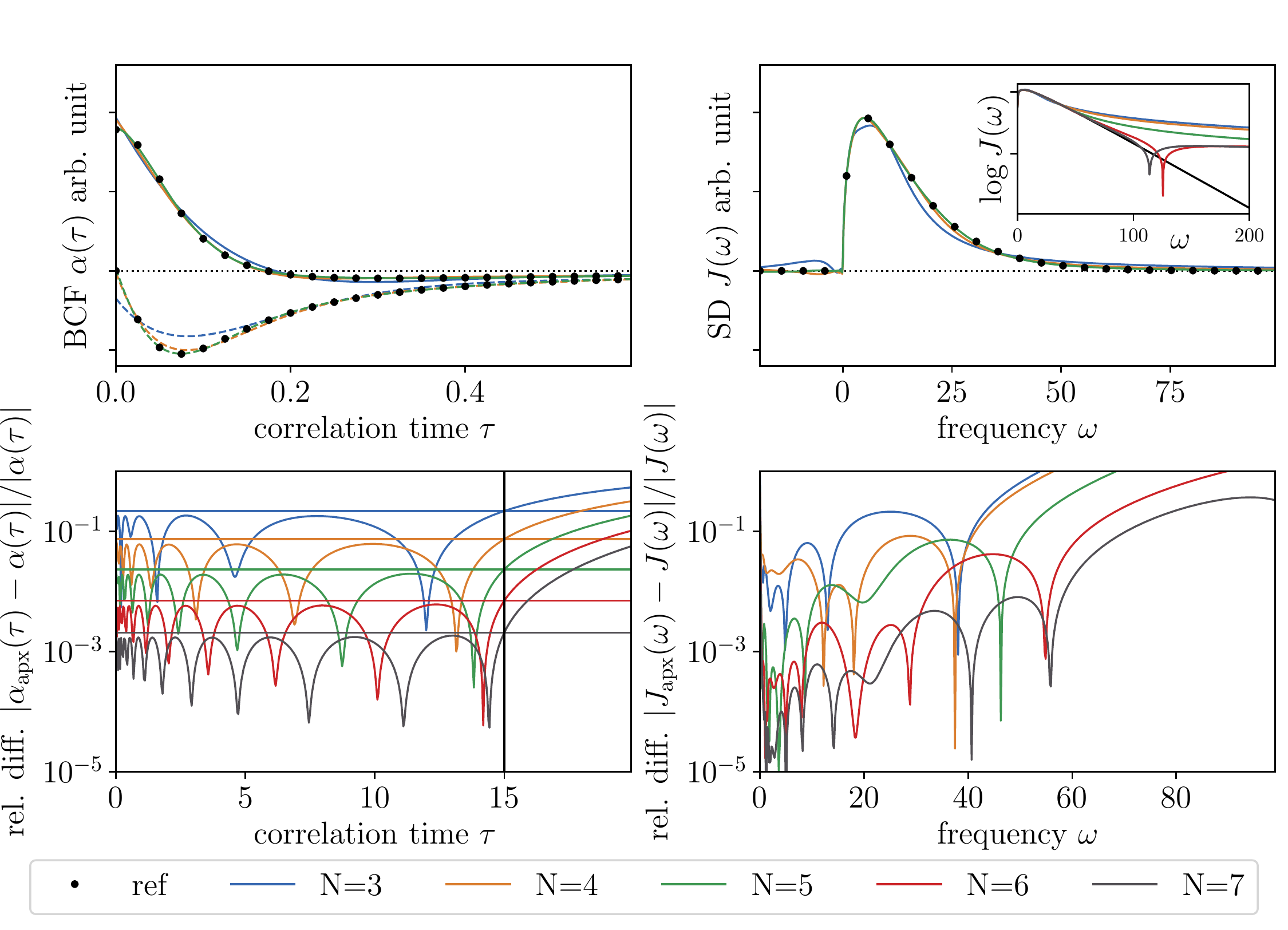}
  \caption{A given sub-Ohmic SD ($s=0.5, \omega_c=10, \alpha=\mathrm{arb.}$, black dots in the upper right panel) yields the BCF at zero temperature shown by the black dots in the upper left panel (real part: solid, imaginary part: dashed). For a given $N$ the minimization procedure of the relative p-norm difference ($p$=10) over the time interval $[0, 15]$ yields an approximative representation of the BCF in terms of a sum of exponentials (colored lines). The ($p=2$) relative difference drawn in the lower left panel shows the scaling of the accuracy with the number of exponentials $N$. The horizontal lines correspond to the maximum relative error of the approximation over the chosen time interval $[0,15]$. The reconstruction of the SD from the approximated BCF is shown by the colored lines in the upper right panel. The logarithm of the SD (see inset) clearly shows that the deviation from the exponential decay shifts to higher frequencies when increasing $N$. From the relative difference of the SD shown in the lower right panel it is seen that an increase in accuracy for the BCF also results in better agreement of the SD over a wide range of frequencies.}
  \label{fig_sd}
\end{figure*}

It turns out that for the class of (sub-/super-) Ohmic SD with exponential cutoff, which is the SD we are concerned about in the application sections,
\begin{equation}
  J(\omega) = \frac{\pi}{2} \alpha \omega_c^{1-s} \omega^s e^{-\frac{\omega}{\omega_c}} \qquad 0 < s
  \label{eqn_sd}
\end{equation}
a straight forward optimization procedure with respect to the BCF parameters $G_j$ and $W_j$ (eq \ref{eqn_bcf}) can be carried out in the time domain to cast the zero temperature BCF to the required form of a sum of exponentials. The accuracy of that representation depends on the number exponentials $N$.
\begin{multline}
  \alpha(\tau) 
  = \il{0}{\infty}{\omega} \frac{J(\omega)}{\pi} e^{-\im \omega \tau} = \frac{\alpha \omega_c^2 \Gamma(s+1)}{2 (1 + \im\omega_c\tau)^{s+1}} \\
  \approx \sum_{j=1}^{N} G_j e^{-W_j \tau} = \alpha_\mathrm{apx}(\tau) \qquad G_i, W_i \in \mathds{C}
  \label{eqn:Ohmic_bcf}
\end{multline}
In order to correctly account for the algebraic decay the optimization procedure is taken to minimize the relative p-norm difference $|\alpha_\mathrm{apx}(\tau) - \alpha(\tau)|_p / |\alpha(\tau)|_p$ over a given time interval. To find good approximations for fixed $N$, the optimization is repeated several times with different random initial parameters $G_j, W_j$. As an example the convergence properties for a sub-Ohmic SD at zero temperature ($s=0.5, \omega_c = 10, t=15$) are shown in Fig. \ref{fig_sd}.

Increasing the time interval for the optimization, namely the time interval where the correct decay of the BCF is guaranteed, requires a larger number of exponentials $N$ to assure the same accuracy. On the other hand for macroscopic environments the BCF decreases in time which suggests that at some point the long time tail might be neglected. Using a fit up to time $t_1$ but propagating HOPS up to time $t_2 > t_1$ means that the eventually non-exponential decay of the BCF is approximated by an exponential behavior for times $t > t_1$. We argue that for a suitably chosen $t_1$ this approximation is fine when examining dynamical properties. Although this scheme might become problematic when dynamical quantities are used to infer spectral properties, especially in the low frequency regime\cite{ritschel_analytic_2014}.

Nonetheless, if one is interested in the dynamics over a given interval of time, a correct representation of the BCF for that particular time interval, including the correct decay behavior, results doubtlessly in the correct dynamics. In contrast, when approximating the SD in order to find an approximation for the BCF, it is not obvious over which frequency range the optimization has to be done.

As a final remark, the straight forward optimization for the parameters $G_j, W_j$ of the BCF approximation is by far not limited to the zero temperature BCF for a (sub-) Ohmic SD with exponential cutoff. We have successfully used the optimization scheme for non-zero temperature BCFs, a regime where HOPS in the form presented here is applicable for Hermitian coupling operators $L = L^\dagger$. Also we see no reason why the fitting of the BCF, and with that HOPS, should not work for different kinds of SDs. However, the zero temperature case was emphasized in this section as we prefer a scheme for HOPS which maps a thermal initial condition to the zero temperature case which will be explained in the following.

\subsection{Non-zero (finite) temperature}

\label{sec_non_zero_remp}

\begin{figure}
  \includegraphics[width=\columnwidth]{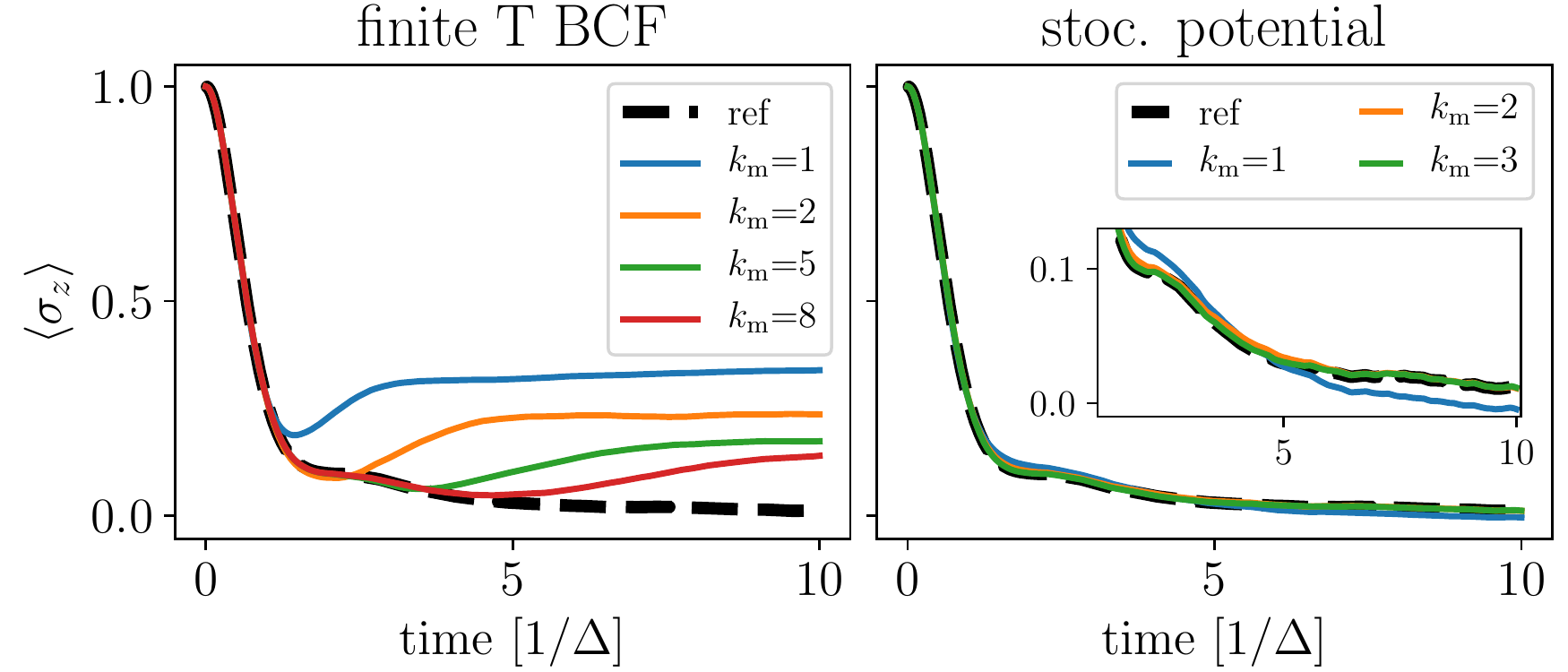}
  \caption{For a self-adjoint coupling operator $L$ thermal initial environmental conditions can be dealt with in HOPS in two ways: by i) employing a fit of the non-zero temperature BCF (left panel) and ii) a stochastic Hermitian contribution to $H_\sys$ in a zero temperature HOPS scheme (right panel). The convergence with respect to the hierarchy depth $k_\mathrm{m}$, directly relating to the computational cost, is shown. 
  The difference is crucial: whereas for the non-zero temperature BCF approach HOPS has not even converged yet at a hierarchy depth $k_\mathrm{m}$=8, the stochastic potential method requires a depth of $k_\mathrm{m}=3$ only. The example presented here is for the spin-boson model of Sec. \ref{sec_chal_regime} for a sub-Ohmic SD ($s$=0.5, $\epsilon$=0, $\omega_c$=10$\Delta$, $\alpha$=0.15, $T$=$\Delta$).}
  \label{fig:temp_method}
\end{figure}

To incorporate thermal initial conditions it turns out that for high temperatures, especially in the strong coupling regime, it is favorable to treat temperature in yet another stochastic manner (see Fig. \ref{fig:temp_method} for a comparison). In this particular representation of the reduced dynamics the zero temperature BCF accounts for the exact quantum mechanical interaction with the bath whereas the effect of non-zero temperature is expressed in terms of a stochastic Hermitian contribution to the system Hamiltonian of the form
\begin{equation}
\begin{gathered}
 H_\sys^\beta(t) = H_\sys + L^\dagger y(t) + L y^\ast(t) \\
 \langle y(t) \rangle = \langle y(t) y(s)\rangle = 0 \qquad \bar n(\beta\omega) = \frac{1}{e^{\beta\omega}-1}\\
 \langle y(t) y^\ast(s)\rangle = \frac{1}{\pi} \il{0}{\infty}{\omega} \bar n(\beta\omega) J(\omega)  e^{-\im\omega (t-s)}
\end{gathered}  
\end{equation}
Similar stochastic potentials have been proposed by unraveling the Feynman-Vernon influence functional\cite{stockburger_<pre>exact_2002, chen_novel_2013, moix_hybrid_2013} where such a Hermitian contribution arises from the real part of the BCF. Our approach however, which is based on the P-function representation for the initial thermal bath state, splits the BCF into the zero temperature and a temperature dependent contribution. For a Hermitian coupling operator $L=L^\dagger$, the stochastic driving $y(t)$ becomes effectively a real valued quantity which can be viewed as a stochastic force $f(t) = 2\mathrm{Re}(y(t))$, like in the Langevin theory, with autocorrelation function
\begin{equation}
  \langle f(t) f(s)\rangle = \frac{2}{\pi} \il{0}{\infty}{\omega} \bar n(\beta\omega) J(\omega) \cos(\omega (t-s))
\end{equation}
which is precisely the temperature dependent contribution to the BCF.
\begin{equation}
\begin{aligned}
  \pi \alpha(\tau) 
  = & \il{0}{\infty}{\omega} J(\omega)\left[\coth(\beta\omega/2)\cos(\omega\tau)-\im\sin(\omega\tau) \right]\\
  = & \il{0}{\infty}{\omega} J(\omega)e^{-\im\omega\tau} + \il{0}{\infty}{\omega} 2\bar n(\beta\omega) J(\omega) \cos(\omega \tau)  
\end{aligned}  
\end{equation}
It should be emphasized that this approach is still exact and fully quantum and can be combined with any open quantum system method that can treat zero temperature initial environmental conditions. In particular, the combination with HOPS is of no major extra computational cost because HOPS is already a stochastic method. There might be a temperature dependence of the required hierarchy depth even for the stochastic temperature method, which is, however, not as crucial as for the non-zero temperature BCF approach. Details of the derivation can be found in Appendix \ref{apx_finite_temp}.

\subsection{Stochastic process generation}

To solve HOPS numerically, the set of differential equations is integrated using standard routines with step size control provided by \texttt{scipy}\cite{jones_scipy:_2017}. The generation of the stochastic processes is done by means of a discrete Fourier Transform method with cubic spline interpolation such that a given tolerance condition is met. The implementation\cite{hartmann_stocproc:_2016} makes use of the fact that the autocorrelation function (BCF) is given by the Fourier Transform of the SD, which can be approximated by the Riemann sum
\begin{equation}
  \alpha(\tau) = \frac{1}{\pi}\il{-\infty}{\infty}{\omega} \tilde J(\omega) e^{-\im\omega\tau} \approx \sum_{k=0}^{n-1} \frac{a_k}{\pi} \tilde J(\omega_k) e^{-\im \omega_k \tau}
\end{equation}
with weights $a_k$ and nodes $\omega_k$. Therefore the stochastic process defined as
\begin{equation}
  z(t) = \sum_{k=0}^{n-1} \sqrt{ \frac{a_k \tilde J(\omega_k)}{\pi}} Y_k e^{-\im \omega_k t}
\end{equation}
with $Y_k$ being complex valued Gaussian distributed random variables with $\langle Y_k \rangle = 0 =  \langle Y_k Y_{k'}\rangle$ and $\langle Y_k Y^\ast_{k'}\rangle = \delta_{k, k'}$, obeys the statistics of the approximated autocorrelation function. Choosing equally distributed nodes $\omega_k = \omega_0 + k \Delta\omega$ and constant weights $a_k = \Delta \omega = (\omega_1 - \omega_0)/(n-1)$, which corresponds to the numeric integration from $\omega_0 - \Delta \omega/2$ up to $\omega_1 + \Delta \omega/2$ with $n$ nodes using the midpoint rule, allows to calculate the time discrete stochastic process using the Fast Fourier Transform (FFT) algorithm: 
\begin{equation}
  z_l = z(t_l) = e^{-\im \omega_0 t_l} \mathrm{FFT}\left( \frac{a_k \tilde J(\omega_k)}{\pi} Y_k \right)
\end{equation}
The time axes is given by $t_l = l\Delta t$ and $n \Delta \omega \Delta t = 2\pi$.
Noting that the interpolated time discrete process yields an autocorrelation function corresponding to the interpolated Riemann sum approximation of the BCF, the error of this method is twofold. On the one hand the interpolation sets a limit on $\Delta t$, whereas the approximation by the Riemann sum requires a small enough $\Delta \omega$, yielding the number of nodes $n$. Both kinds of tolerance are specified by means of the maximum absolute difference. Choosing the absolute difference over the relative difference is motivated by the fact that the autocorrelation function, calculated in practice by drawing samples, will have noise imposed which scales with one over square root of the samples. This in turn means that the decay of the autocorrelation function beyond the noise level cannot be observed. Roughly speaking, a highly accurate (low noise level) calculation using HOPS requires many samples which in turn requires a lower tolerance for the stochastic process generation.

\section{Benchmarking HOPS with the spin-boson Model}

In the following the spin-boson model is considered which, despite its simple form, has been in the focus of open quantum system research for decades\cite{leggett_dynamics_1987}. In terms of the HOPS method it means that the system Hamiltonian $H_\sys$ and the coupling operator $L$ are specified to be
\begin{equation}
  H_\sys = \epsilon \sigma_z + \Delta \sigma_x \qquad L = L^\dagger = \sigma_z
\end{equation}
Furthermore the environment is assumed to be of (sub-) Ohmic structure. The corresponding BCF (eq \ref{eqn:Ohmic_bcf}), which enters the HOPS method, is parametrized by $s$, the low frequency power law behavior of the SD, $\omega_c$, the cut of frequency, and $\alpha$, the coupling strength.

We will now compare our results for the spin-boson model with other methods namely (i) the quantum optical master equation and an extension correctly accounting for the sub-Ohmic case, and (ii) HEOM and eHEOM in the high temperature limit which suggests to use a purely classical treatment of the environment as done in Ref.\cite{tang_extended_2015}. In the last Section we compare with (iii) the ML-MCTDH method at zero temperature for which in the strong coupling regime the numerical effort becomes significant. To go beyond that we employ HOPS to also solve for non-zero temperature at approximately the same numerical cost.

\subsection{The quantum optical master equation}

First the comparison of the exact population dynamics gained from HOPS with the dynamics calculated using a Lindblad master equation is drawn. This master equation, which reads
\begin{multline}
 \ddt \rho_\sys(t) = -\im [H_\sys + H_\mathrm{lamb}, \rho_\sys(t)] \\
   + \sum_{\omega} \tilde J(\omega) (2 L_\omega \rho_\sys(t) L^\dagger_\omega - [L^\dagger_\omega L_\omega, \rho_\sys(t)]_+)
\label{eqn:ME}
\end{multline}
was obtained from the microscopic model by applying a Born, Markov and rotating wave approximation\cite{breuer_theory_2007} which obviously limits its validity to a quite special regime.

In the standard Markovian limit, using $\Gamma(t, \omega) := \il{0}{t}{s}\alpha(t-s)e^{\im \omega s}$, the so called Lamb shift contribution is scaled by the imaginary part  of
\begin{equation}
  \tilde J(\omega) + \im S(\omega) := \lim_{t \rightarrow \infty} \Gamma(t, \omega) = \il{0}{\infty}{s} \alpha(t-s)e^{\im \omega s} 
  \label{eqn:gamma}
\end{equation}
and reads $H_\mathrm{lamb} = \sum_{\omega} S(\omega) L^\dagger_\omega L_\omega$. When denoting the eigenvalues of $H_\sys$ by $\pm \lambda = \pm\sqrt{\epsilon^2 + \Delta^2}$, the index $\omega$, which corresponds to all possible differences of eigenvalues, can take the values $-2\lambda, 0$ and $2\lambda$. The components $L_\omega$ of the decomposition of the coupling operator $L$ take the form:
\begin{equation}
\begin{gathered}
  L_0 = \cos(2\theta) (\cos(2\theta) \sigma_z + \sin(2\theta)\sigma_x) \\
  L_{2\lambda} = \sin(2\theta)\left(\frac{\sin(2\theta)}{2}\sigma_z + \sin^2(\theta) \sigma_+ - \cos^2(\theta)\sigma_-\right) \\
  L_{-2\lambda} = L^\dagger_{2\lambda} \qquad \tan(2\theta) = \frac{\Delta}{\epsilon}
\end{gathered}  
\end{equation}
In order to write the non-zero temperature BCF as a Fourier transform it is convenient to introduce the pseudo SD $\tilde J(\omega) = \frac{J(\omega)}{1 - \exp(-\beta\omega)}$ with also negative frequency contributions where $J(-\omega) := -J(\omega)$.

Its behavior at $\omega$=0, which contributes to the damping terms in the master equation, depends on the parameter $s$, determining the low frequency limit of $J(\omega)$. In case of an Ohmic SD ($s$=1) the limit exists and takes the value $\tilde J(0)=\pi\alpha/2/\beta$. The same holds true for super-Ohmic SD ($s>1$) yielding $\tilde J(0)=0$, whereas for the sub-Ohmic case ($s<1$) the pseudo SD diverges. This behavior indicates that the dynamics obtained from the standard master equation (\ref{eqn:ME}) depends discontinuously on the parameter $s$ which contradicts the plausible argument that an infinitesimal change in the environmental parameters should only yield an infinitesimal change in the dynamical properties. Note that the discontinuity becomes evident only in case of non-zero bias ($\epsilon>0$). Otherwise $L_0$ is identical zero which suppresses the contribution of $\tilde J(0)$. 

\begin{figure}
  \includegraphics[width=\columnwidth]{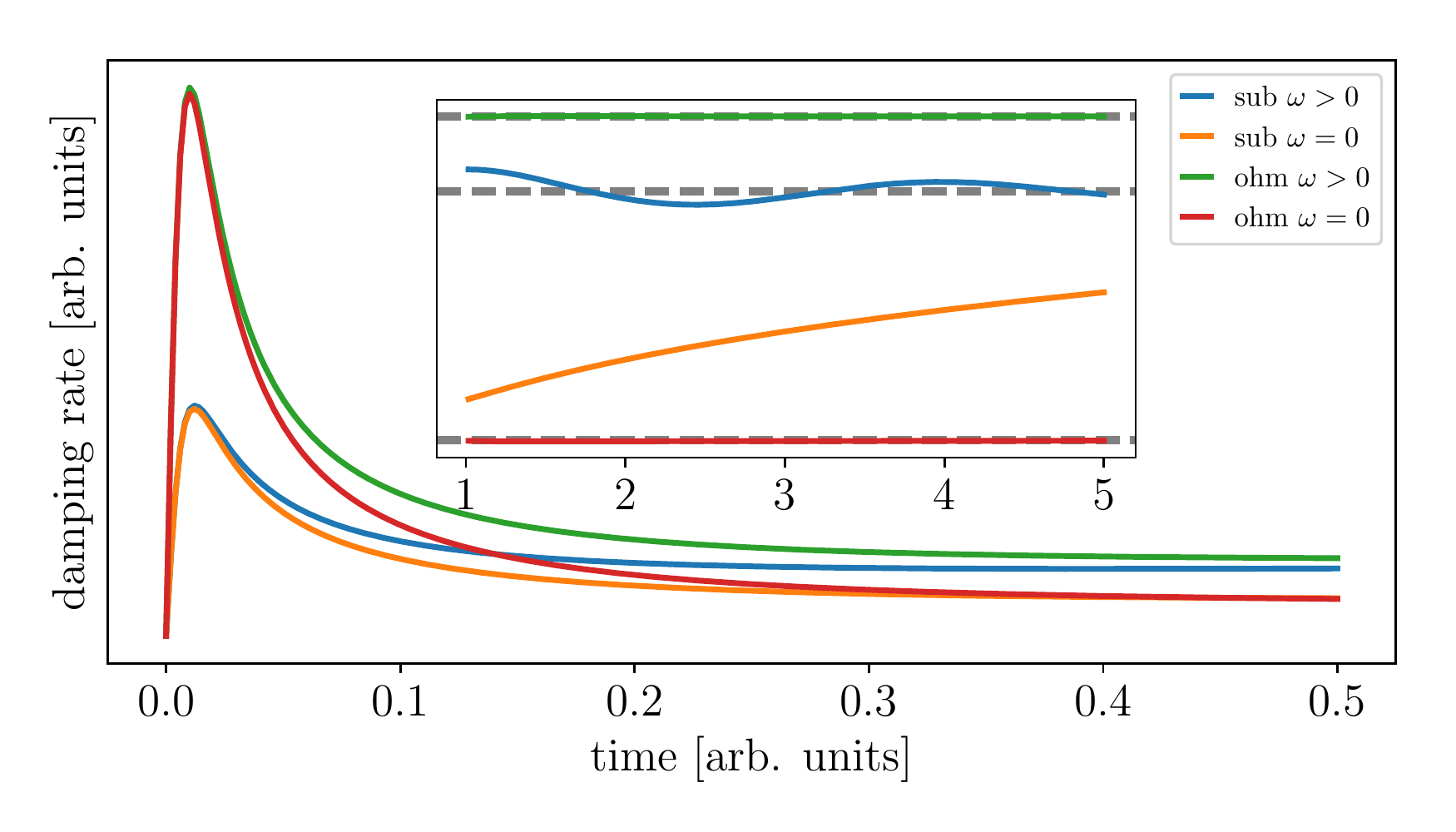}
  \caption{To obtain time independent damping rates for the master equation, the decay of the BCF is assumed to be the fastest time scale such that the integral $\Gamma(t,\omega) = \mathrm{Re}\big\{ \il{0}{t}{s} \alpha(s) e^{\im\omega s} \big\}$ reaches its stationary value much faster than the typical time scale of the dynamics induced by the interaction. The plot shows that the assumption may hold true in the Ohmic case for both $\omega>0$ (green line) and $\omega$=0 (red line). By contrast, in the sub-Ohmic case there is no stationary value for $\omega$=0 (yellow line). So replacing $\Gamma(t)$ by $\Gamma(\infty) = \infty$ yields an unacceptable approximation. The dashed lines in the inset correspond to the stationary value $\Gamma(t=\infty,\omega) = \tilde J(\omega)$, if existing.}
  \label{fig:markov_apprx}
\end{figure}

It has been pointed out\cite{noh_<pre>consistent_2014} that these problems arise from taking the limit $t\rightarrow\infty$ in (\ref{eqn:gamma}). Keeping the actual $\Gamma(t, \omega)$ yields time dependent coefficients for the master equation, restoring a growing but finite value for the $\omega$=0 contribution (see Fig. \ref{fig:markov_apprx})
which is consistent with other results\cite{kehrein_spin-boson_1996, shnirman_dephasing_2003}. The usefulness of the consistent perturbative treatment proposed in\cite{noh_<pre>consistent_2014} is confirmed by finding very good agreement with our exact numerical results using HOPS (see Fig. \ref{fig:bm_sub}).

\paragraph{The Ohmic case $s=1$:}

\begin{figure}
  \includegraphics[width=\columnwidth]{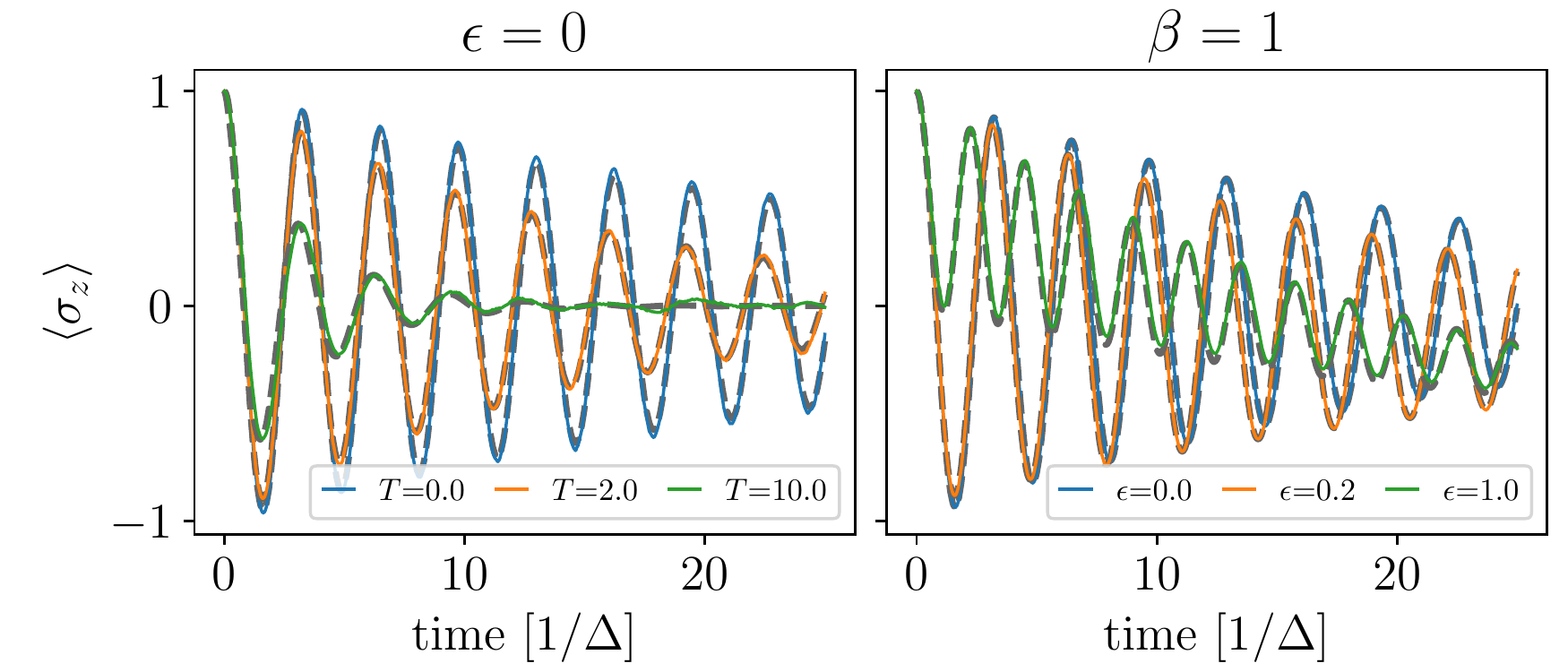}
  \caption{In case of small coupling ($\alpha$=0.01), large cutoff frequency ($\omega_c$=100$\Delta$) and an Ohmic SD ($s$=1) the dynamics for the spin-boson model gained from HOPS (colored lines) matches very well the results calculated using the usual quantum optical master equation (dashed gray lines) with constant rates. The left panel shows the unbiased case for various temperatures measured in units of $\Delta$. The dependence on the bias in units of $\Delta$ for a fixed temperature $T$=$\Delta$ is shown in the right panel.}
  \label{fig:bm_Ohmic}
\end{figure}

\begin{figure}
  \includegraphics[width=\columnwidth]{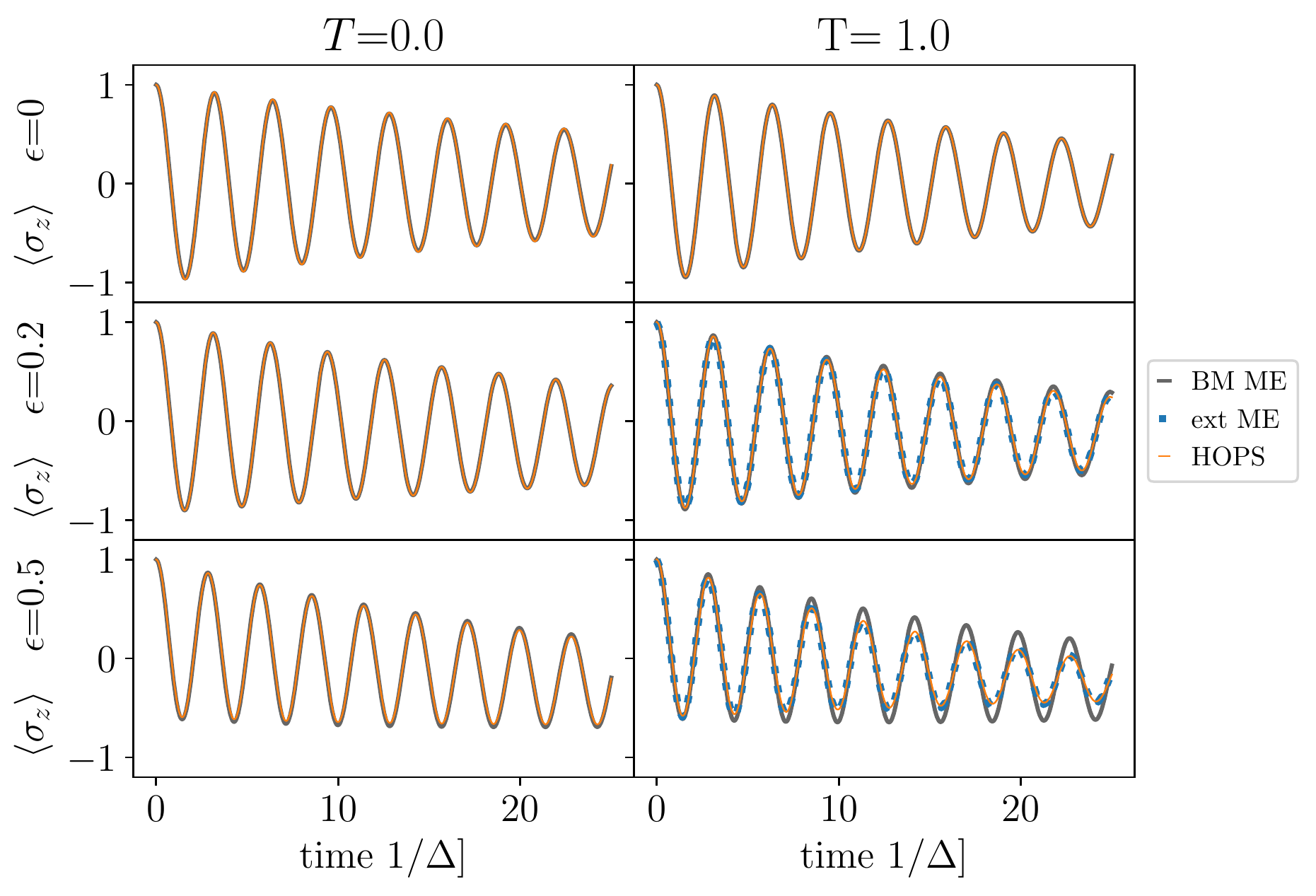}
  \caption{The spin-boson model with a sub-Ohmic SD with parameters $\alpha$=0.01$\omega_c^{s-1}$, $\omega_c$=100$\Delta$ and $s$=0.8 is considered. As expected, in the unbiased case $\epsilon$=0 (upper panels) perfect agreement between the quantum optical (Born Markov) master equation with skipped $\omega$=0 contribution (BM ME, gray line) and HOPS (orange line) can be seen for zero and non-zero temperature. Increasing the bias $\epsilon$ (middle and lower panels) reveals the influence of the $\omega$=0 contribution. For zero temperature the effect exists but is not visible in this plot. When visible (mid right and lower right panel) HOPS perfectly matches the dynamics from the consistent perturbative master equation (ext ME) with a special treatment of diverging $\tilde J(0)$ contribution.}
  \label{fig:bm_sub}
\end{figure}

Choosing a small coupling parameter $\alpha$=0.01 and a large cutoff frequency $\omega_c$=100$\Delta$ (eq \ref{eqn_sd}), which should justify the Born and Markov approximation, and fixing the SD to the Ohmic case, we find agreement between the dynamics gained from the master equation and HOPS (see Fig. \ref{fig:bm_Ohmic}). Although the Ohmic case has been studied excessively in contrast to the more challenging sub-Ohmic regime we still show the Ohmic case to test the procedure of fitting the BCF in time which does not distinguish between the Ohmic and sub-Ohmic case. Therefore agreement with other methods in the Ohmic case provides evidence that HOPS should also work in the sub-Ohmic and any other case as long as the fitting works well.

As expected increasing the coupling strength or lowering the cutoff frequency significantly will result in disagreement of the two methods. Such graphs are not shown.

For completeness, we have used $N=6$ exponential summands to fit the BCF in the interval $[0, 0.5/\Delta]$. At $t=0.5/\Delta$ the BCF has decreased by three orders of magnitude $|\alpha(0.5/\Delta)| / |\alpha(0)| < 10^{-3}$. The maximum relative error with respect to that interval is less than $4\cdot 10^{-3}$. The reduced state was calculated by averaging over $10^4$ stochastic trajectories.

\paragraph{The sub-Ohmic case $s<1$:}

To compare HOPS with the Born-Markov results in the sub-Ohmic case, special care has to be taken. As pointed out earlier the pseudo SD diverges at $w$=0 seemingly yielding an infinite $\tilde J(0)$ contribution. In a more careful perturbative treatment (Ref.\cite{noh_<pre>consistent_2014}) this contribution is replaced by the time dependent rate $\mathrm{Re} (\Gamma(t, \omega=0)) $. 
For all parameters very good agreement between HOPS and this extended master equation can be seen in Fig. \ref{fig:bm_sub}. This provides a first consistency check that HOPS can deal with sub-Ohmic environments.

It is worth pointing out that the special treatment of the pseudo SD contribution at $w$=0 is not limited to the spin-boson model but should be considered rather general when dealing with a master equation of the kind of eq \ref{eqn:ME} in combination with sub-Ohmic environments.

As in the Ohmic case we used the same parameters for fitting the BCF yielding a slightly higher maximum relative difference of less than $7\cdot 10^{-3}$. As before, $10^4$ samples where used to obtain the reduced dynamics.

\subsection{HEOM, eHEOM and classical noise}

Next we consider a regime where presumably the environmental influences can be accounted for by stochastic forces alone. This is possible whenever the imaginary part of the BCF can be neglected with respect to the real part and
requires very high temperatures. In that limit the exact reduced dynamics can be obtained via a stochastic Hamiltonian (stoch. Ham.) method\cite{tang_extended_2015}. When comparing with that exact method, Tang et al.\cite{tang_extended_2015} pointed out that the usual HEOM with a Meier-Tannor (MT) decomposition of the SD (MT HEOM) is not entirely suitable for sub-ohmic environments. However, the eHEOM\cite{tang_extended_2015} approach, an extension of HEOM not relying on a BCF of the form of a sum of exponentials, cured the problem. It should be pointed out that due to the MT-decomposition for HEOM a small imaginary contribution is included, whereas the eHEOM method relies only on a real valued function basis set to decompose the BCF, therefore neglects the imaginary part of the BCF by construction.

In the following we compare HOPS with these three methods. Regardless of the temperature the HOPS calculation is carried out in the fully quantum regime including the imaginary part of the BCF. All data except HOPS were taken from Ref.\cite{tang_extended_2015}. 

\begin{figure}
  \includegraphics[width=\columnwidth]{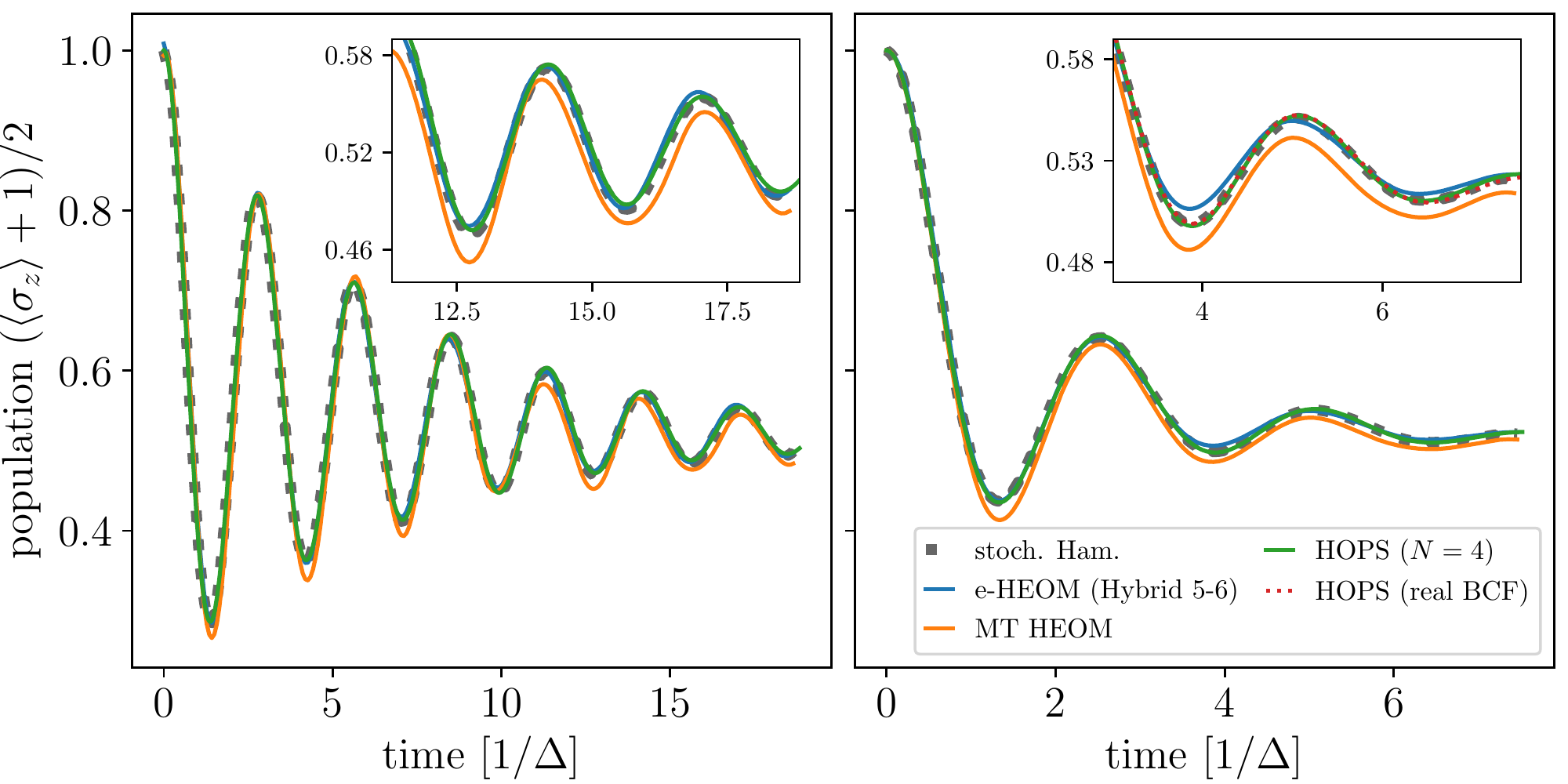}
  \caption{In the high temperature limit the imaginary part of the BCF is assumed to be negligible allowing for a treatment of the environmental influences via stochastic forces (stoc. Ham. method). For the unbiased ($\epsilon$=0) spin-boson model in that regime, HOPS is compared with HEOM and eHEOM. 
  For the left panel ($\alpha = 0.266$, $\omega_c = 0.531\Delta$, $T = 2.09\Delta$, $s$=0.5) eHEOM and HOPS match nicely the reference data from the stoch. Ham. method, whereas the usual HEOM yields slightly different dynamics as pointed out in Ref. \cite{tang_extended_2015}. In the right panel ($\alpha = 0.106$, $\omega_c = 1.33 \Delta$, $T = 5.21\Delta$, $s$=0.5) the zoomed in view of the inset reveals that HOPS really closely matches the reference, whereas eHEOM slightly deviates concluding that the eHEOM method might need more basis functions to approximate the BCF or a higher hierarchy depth. For consistency we have also run HOPS by fitting the non-zero temperature real valued BCF (red dots) which matches the reference data just as HOPS does. The numeric values of the used parameters (specified in units of $\Delta$) are given with accuracy of three digits and corresponds to the parameters given in Fig. 5d and Fig. 5b. of Ref. \cite{tang_extended_2015}}
  \label{fig:extHEOM}
\end{figure}

As seen in Fig. \ref{fig:extHEOM} and pointed out by Tang et al.\cite{tang_extended_2015} the MT decomposition of a sub-Ohmic SD yields inaccurate results (yellow line) which can be improved by making use of the extended HEOM (blue line). The same improvement can be achieved by employing HOPS with a fit of the BCF in the time domain. In contrast, the e-HEOM results made use of a decomposition of the BCF requiring 31 orthogonal functions and a hierarchy depth of 6 which results in a set of 31.675.182 coupled differential equations, whereas for HOPS the fit of the BCF required 4 exponential terms to achieve an accuracy (relative difference) of less than 1\% and a hierarchy depth of 5 is more than sufficient yielding a set of 252 coupled differential equations.

Furthermore considering the parameter set corresponding to the right panel in Fig. \ref{fig:extHEOM} a small discrepancy between e-HEOM and the exact stochastic Hamiltonian method is observed. However, the fully quantum mechanical calculation using HOPS matches the dynamics gained from the stochastic Hamiltonian method very well supporting the validity of neglecting the imaginary part of the BCF. To check consistency we have also set up HOPS for the real valued non-zero temperature BCF (red dotted line in the right panel of Fig. \ref{fig:extHEOM}) $\alpha_\mathrm{re}(\tau) = \frac{1}{\pi}\il{0}{\infty}{\omega} J(\omega) \coth(\beta\omega/2)\cos(\omega\tau)$ by fitting this particular BCF which reproduces the dynamics from the stochastic Hamiltonian method as well. As a reminder the fully quantum mechanical calculation via HOPS involved fitting the zero temperature BCF and includes temperature in a stochastic manner similar to the stochastic Hamiltonian method.

These results show that HOPS, which operates in the fully quantum regime, is very well capable of dealing with high temperature environments with predominant classical influence (thermal fluctuations) on the quantum system. From the way how temperature was included in HOPS, this was expected (see Sec. \ref{sec_non_zero_remp}).

\begin{figure}
  \includegraphics[width=\columnwidth]{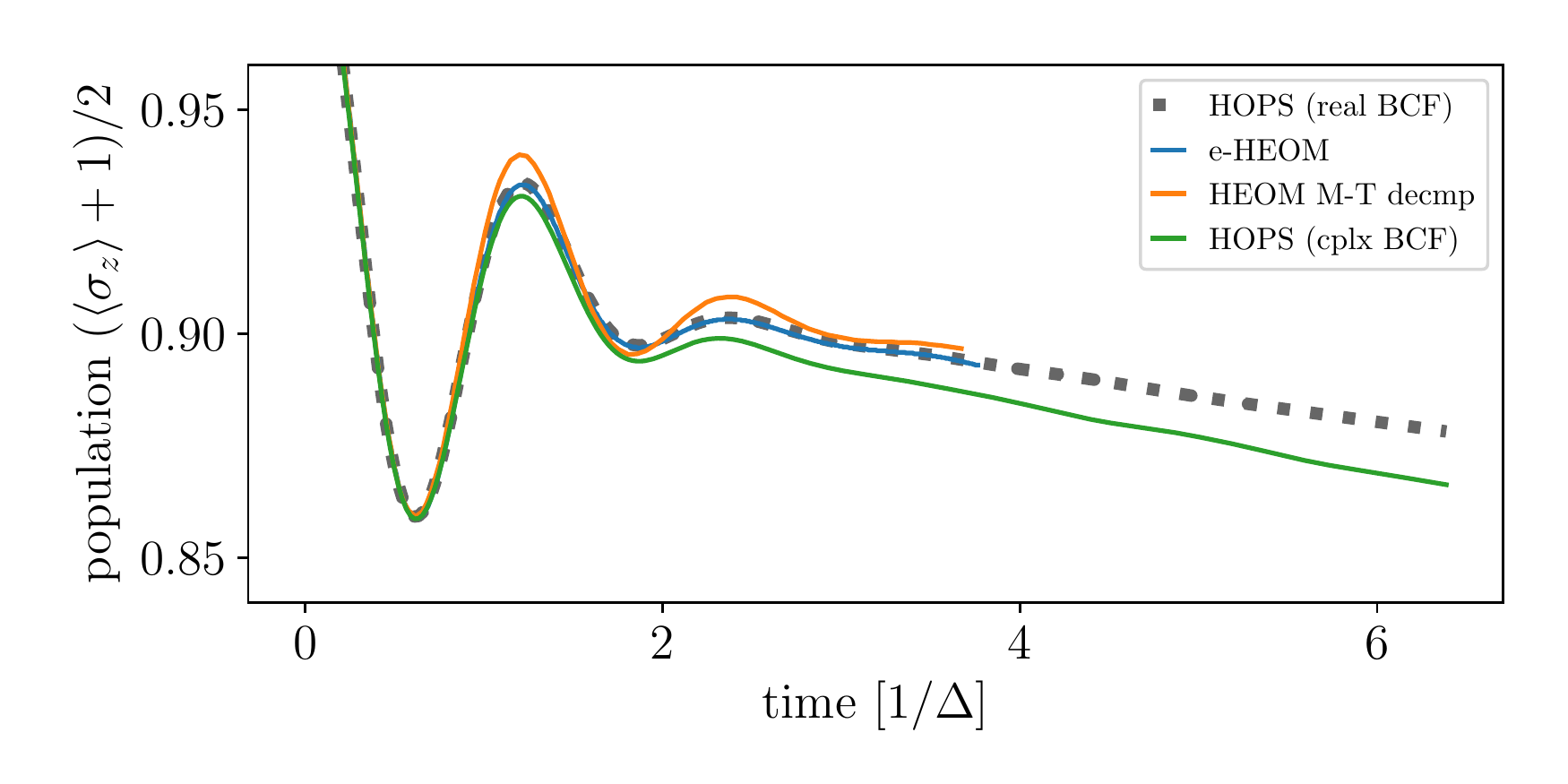}
  \caption{The biased ($\epsilon=2.5\Delta$) spin-boson model is considered with parameters $\alpha = 0.0106$, $\omega_c = 1.65\Delta$, $T = 10.4\Delta$, $s$=0.5 corresponding to the parameters from Ref. \cite{tang_extended_2015} Fig. 6a. The population dynamics gained from the fully quantum mechanical calculation using HOPS (green line) decays faster than the dynamics resulting from the other methods neglecting the imaginary part of the BCF. This example shows that it is not straight forward to tell a priory whether the imaginary part of the BCF can be neglected or not.}
  \label{fig:extHEOM_fig6}
\end{figure}

Additionally we investigated a different set of parameters with a dominating bias $\epsilon$, weak coupling and high temperature (see Fig. \ref{fig:extHEOM_fig6}). The parameters are again taken from Ref.\cite{tang_extended_2015} Fig. 6a where they serve as an example for the biased spin-boson model. In contrast to the set of parameters discussed before we find that the quantum nature of the bath may not be neglected. In Fig. \ref{fig:extHEOM_fig6} we show that the dynamics calculated using HOPS with real valued BCF (dashed gray line) coincides well with the results from e-HEOM (blue line) and the stochastic Hamiltonian method (not shown). The usual HEOM with MT decomposition (yellow line) gives slightly higher oscillations. However, invoking HOPS with the true BCF (green line) yields the same oscillatory behavior as HOPS with the real valued BCF but with a slightly faster decay. In conclusion the quantum nature of the bath resulting in a faster decay may not be neglected for this particular set of parameters.

\FloatBarrier

\section{HOPS for the challenging parameter regime}

\label{sec_chal_regime}

Now we turn to the scenario where an unbiased qubit ($\epsilon$=0) is strongly coupled to a sub-Ohmic bath ($s$=0.5, $\omega_c$=10$\Delta$) which becomes numerically demanding when increasing the coupling strength. The zero temperature case has been studied by Wang and Thoss\cite{wang_coherent_2010} using multilayer multiconfiguration time-dependent Hartree (ML-MCTDH) method. We reproduce the ML-MCTDH results and provide additional data for thermal initial conditions.

In order to apply HOPS for that regime several special aspects of the method should be pointed out again. 

\begin{figure}[b]
  \includegraphics[width=\columnwidth]{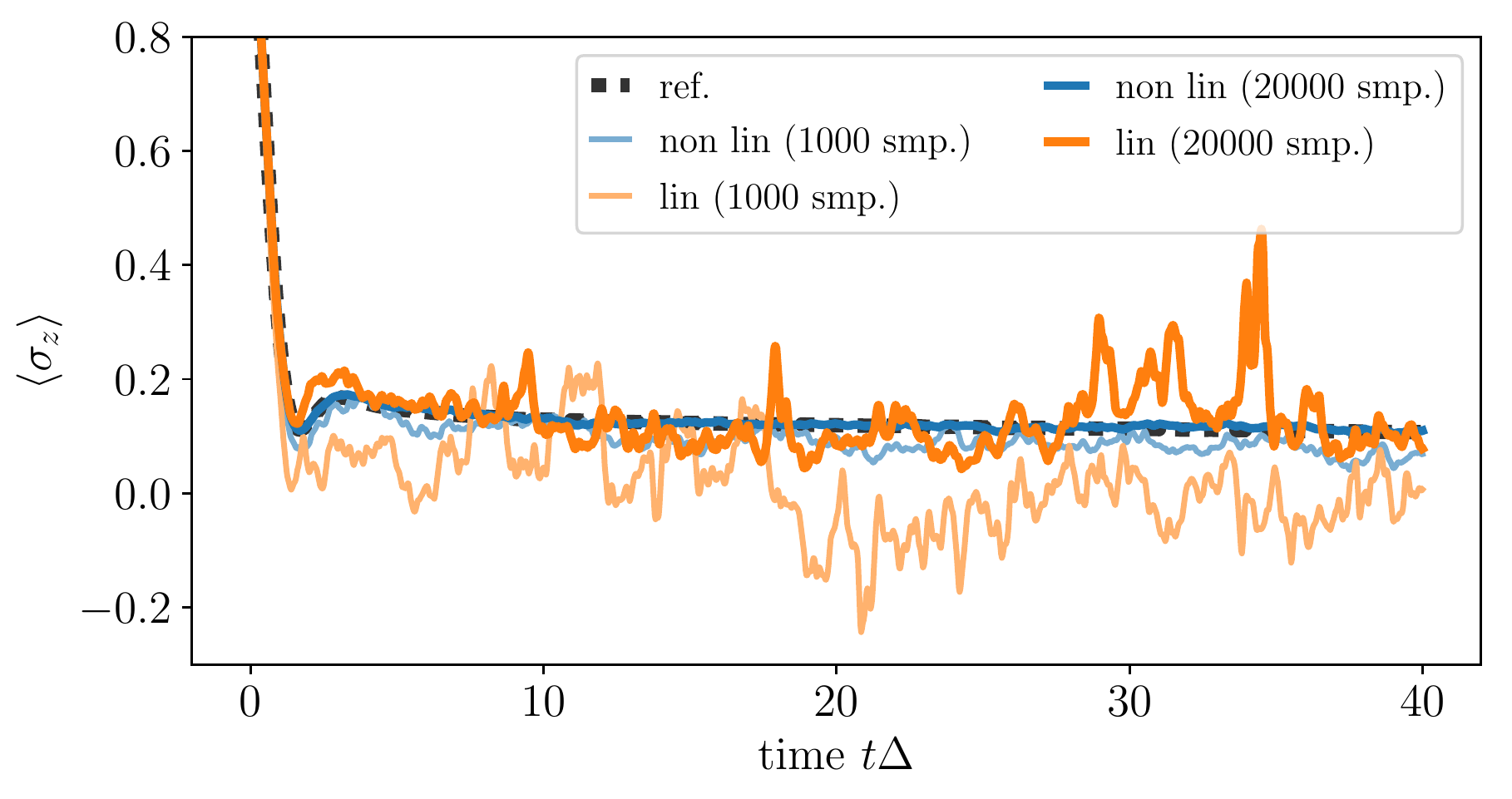}
  \caption{Comparison of the convergence properties for the linear HOPS (orange line) and the non-linear HOPS (blue line) in the strong coupling regime ($\alpha$=0.2, $s$=0.5, $\omega_c$=10$\Delta$, $\epsilon=0$, $T$=0). As reference (gray dashed line) the non-linear HOPS method with 100.000 samples was used.}
  \label{fig:lin_vs_nonlin}
\end{figure}

\paragraph{The non-linear} NMQSD equation \cite{strunz_open_1999} which results in a non-linear variant of the HOPS method \cite{suess_hierarchy_2014} (see end of Sec. \ref{sec_HOPS_derivation}) is inevitable for the strong coupling regime. In Fig. \ref{fig:lin_vs_nonlin} it is clearly seen that the noise of the $\langle \sigma_z \rangle$ dynamics is not only larger for the linear HOPS compared to the non-linear HOPS, but also does not seem to decrease when increasing the number of samples by a factor of 20. In contrast, when applying the non-linear version the noise level decreases notably yielding the converged exact dynamics.

Further we found that for a fixed hierarchy depth, the extra numerical effort for solving the non-linear set of differential equations is negligible compared to its linear version. Additionally we could confirm that for small coupling parameters the linear and the non-linear version perform about the same\cite{de_vega_non-markovian_2005}, here in particular concerning the required hierarchy depth. Note that in the strong coupling regime the comparison of the required hierarchy depth between the linear and non-linear hierarchy is not meaningful as the linear version is not applicable. We therefore consider the non-linear hierarchy as being the general method of choice for practical applications.

\begin{figure*}
  \includegraphics[width=\graphwidth]{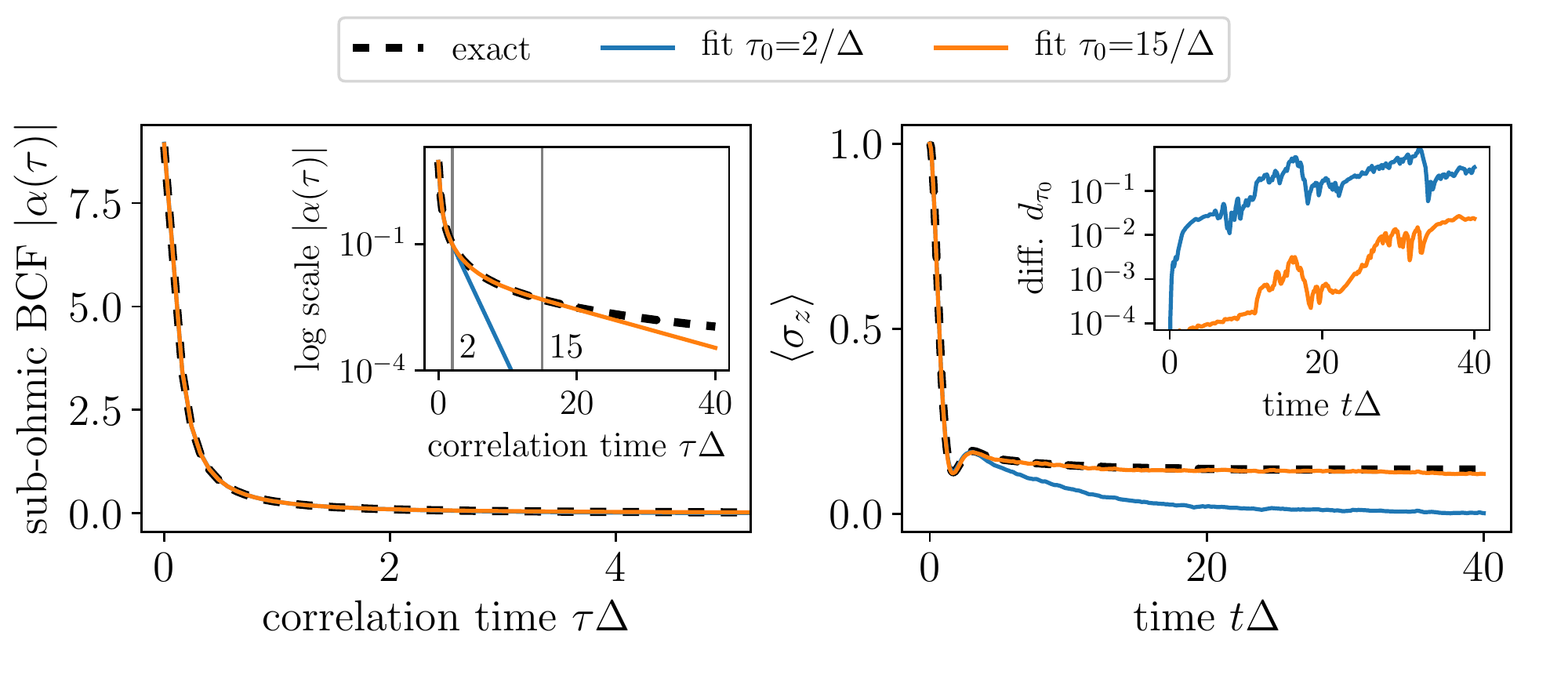}
  \caption{The influence of the algebraic decay of the sub-ohmic BCF on the spin-boson model ($\alpha$=0.2, $s$=0.5, $\omega_c$=10$\Delta$, $\epsilon$=0, $T$=0) is shown. In the non-logarithmic plot of the absolute value of the BCF (left panel) the decay beyond $\tau_0 = 2/\Delta$ is not visible. However the logarithmic plot (inset) reveals the significant difference which is reflected in the $\langle \sigma_z \rangle$ dynamics shown in the right panel. The inset of the right panel shows the absolute difference $d_{\tau_0}$ between a single stochastic state vector for various fits.}
  \label{fig:short_fit_bcf}
\end{figure*}

\paragraph{The ``slow'' algebraic decay} of the sub-Ohmic BCF needs to correctly be accounted for when approximating the BCF in terms of exponentials. From the exponential form of the approximation it is clear that the long time behavior of the fit will be of exponential kind where the start of the exponential decay is determined by the end $\tau_0$ of the fit interval $[0, \tau_0]$. Fitting the BCF up to time $\tau_0 = 2/\Delta$ is motivated by the fact that at $2/\Delta$ the BCF has decayed by about two orders of magnitude. Differences for larger correlation times between the fit and the exact BCF can not be resolved by looking at the absolute value of the BCF (see Fig. \ref{fig:short_fit_bcf} left panel) which suggests the faulty conclusion that the fit is perfectly valid also for larger correlation times. However the logarithmic plot of the absolute value of the BCF (see inset of the left panel) reveals the two distinct decay kinds. Additionally, a fit up to $\tau_0 = 15/\Delta$ is considered. 

The difference in the dynamics based on these two fits is clearly visible in the right panel of Fig. \ref{fig:short_fit_bcf}. Whereas the fit with $\tau_0 = 2/\Delta$ tends to zero for longer times, the more accurate fit matches very well the reference data obtained by Wang et al. \cite{wang_coherent_2010} via ML-MCTDH method. Note that the fit of the BCF with $\tau_0 = 15/\Delta$ may very well be used to propagate HOPS up to $t=40/\Delta$. In order to verify that the correction due to an even better fit up to $\tau_0 = 40/\Delta$ is minor, it is sufficient to look at the convergence, with respect to the fit, of a single stochastic state vector (zeroth order of the hierarchy, fixed stochastic process $z^\ast_t$). Since in the non-linear HOPS method the reduced state is reconstructed from the normalized stochastic trajectories the correction is estimated from the convergence of these normalized stochastic trajectories $\tilde \psi := \psi[z^\ast]/|\psi[z^\ast]|$. Using the fit with $\tau_0 = 40/\Delta$ as reference the difference $d_{\tau_0} = |\tilde \psi_{\mathrm{fit}\;\tau_0} - \tilde \psi_{\mathrm{fit}\;40/\Delta}|$ reflects the error due to the exponential decay starting at $\tau_0$. It is seen in the inset of the right panel of Fig. \ref{fig:short_fit_bcf} that the naive fit with $\tau_0 = 2/\Delta$ results in a significantly different stochastic state vector, explaining the deviation in the reduced dynamics. However, the difference for the fit with $\tau_0 = 15/\Delta$ is of the order of 1\% which is consistent with the very good agreement of the dynamics gained from HOPS (BCF fit up to $\tau_0 = 15/\Delta$) with ML-MCTDH.

It should be pointed out that increasing $\tau_0$ while keeping the accuracy of the fit at the same level results in an increase of the number of exponential terms $N$ which is drastically reflected in the number of auxiliary states. To provide an example, in order to achieve a maximum relative difference of about $10^{-3}$ over the interval $[0, \tau_0]$ the approximation requires $N=5$ for $\tau_0 = 2/\Delta$, $N=8$ for $\tau_0 = 15/\Delta$ and $N=10$ for $\tau_0 = 40/\Delta$. For a fixed hierarchy depth of 10 this results in 3002, 43.757, 352.715 auxiliary states respectively.

\paragraph{Many samples and a large hierarchy depth} are required for the strong coupling regime. The coupling strength $\alpha$ simply scales the BCF which enters HOPS in two ways.

First it enters via the autocorrelation function of the stochastic process $z^\ast_t$ which means that the amplitude of $z^\ast_t$ scales with $\sqrt{\alpha}$. This results in an increase of the noise for each stochastic state vector with the coupling strength, which in turn requires more samples to reach for the same smoothness of the reduced state. Notably when averaging over normalized stochastic state vectors, as in the non-linear HOPS, the distribution of the stochastic state vectors is bound independently of the coupling strength which results in a coupling strength independent standard deviation of the average proportional to $1/\sqrt{N}$. To obtain smooth results in the parameter regime presented here $10^5$ up to $10^6$ samples where used.

\begin{figure*}
  \includegraphics[width=\graphwidth]{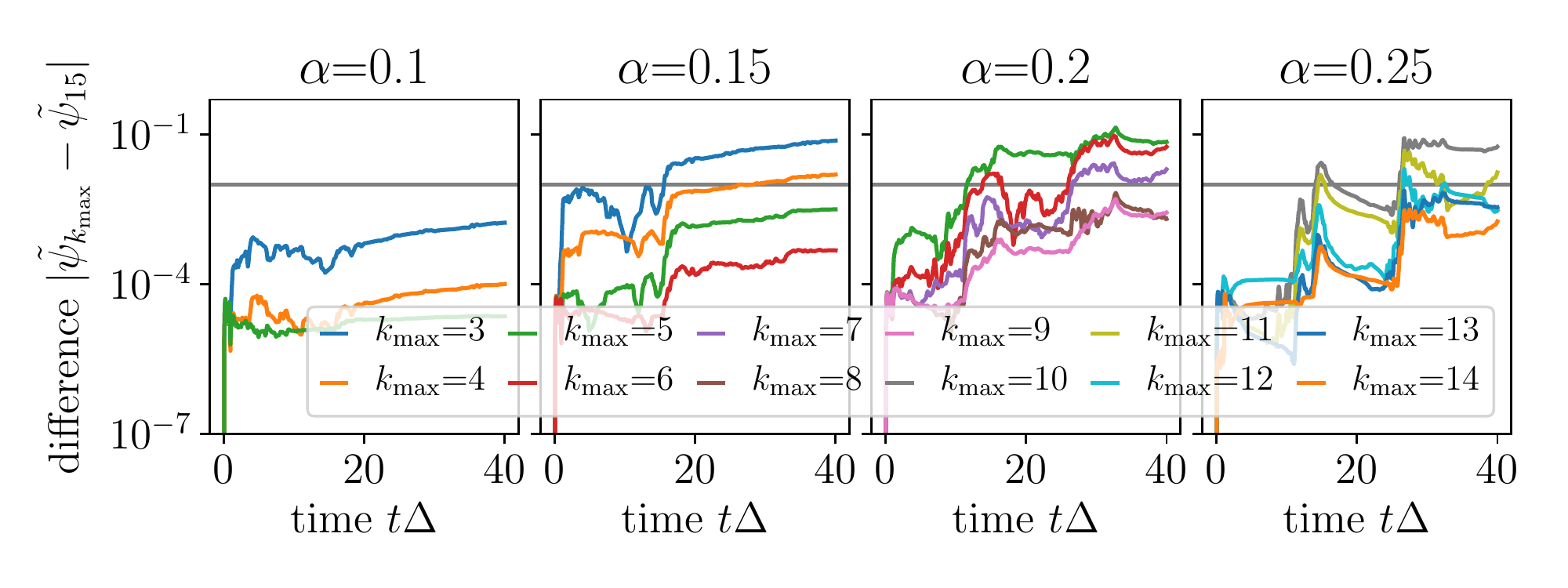}
  \caption{For a fixed stochastic process $z^\ast_t$ HOPS was run for various cutoff levels $k_\mathrm{max}$. The convergence of the stochastic state vector with respect to the hierarchy depth can be seen from the absolute difference between the normalized stochastic state vector with hierarchy depth $k_\mathrm{max}$ and a reference depth of 15. The hierarchy depth was chosen such that the difference stays below a given threshold of $10^{-2}$. The data shows that increasing the coupling strength requires a significant increase of the hierarchy depth.}
  \label{fig:kmax_conv}
\end{figure*}

Second, the BCF enters via the approximation in terms of the sum of exponentials where the coupling strength $\alpha$ directly scales the fit parameters $G_j$ which in turn are responsible for the coupling of different auxiliary states. A larger coupling results in a population of larger hierarchy levels, which may influence the dynamics of the stochastic state vector, requiring a larger hierarchy cutoff level (see Fig. \ref{fig:kmax_conv}).

\paragraph{A special treatment of thermal initial conditions} is necessary for HOPS to converge with respect to the hierarchy depth which has already been discussed in Sec. \ref{sec_non_zero_remp} (see also Fig. \ref{fig:temp_method}): we include thermal fluctuations as Hermitian contribution to the system Hamiltonian. A rigorous derivation for this approach can be found in Appendix \ref{apx_finite_temp}.

\begin{figure*}
  \includegraphics[width=\graphwidth]{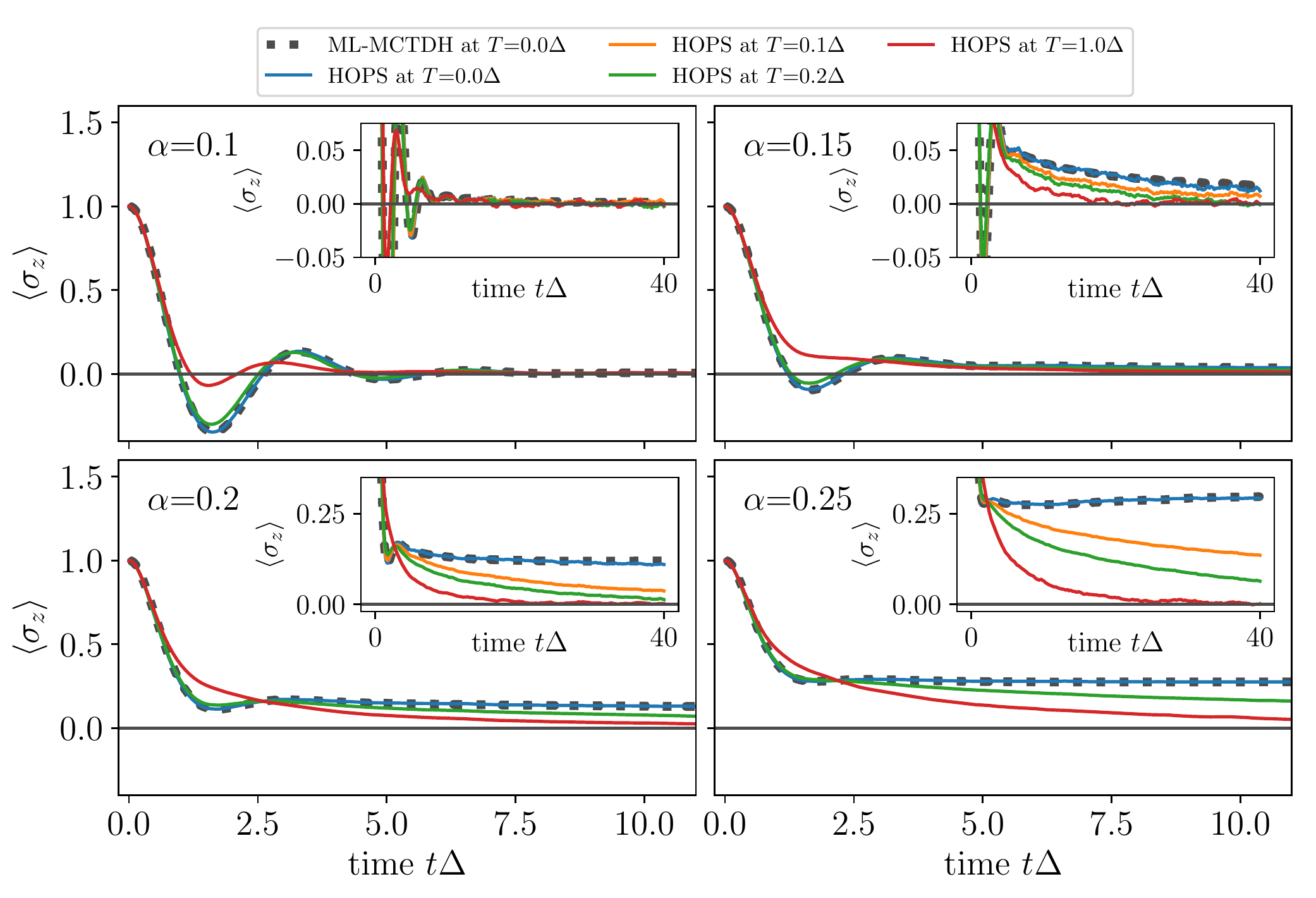}
  \caption{The dynamics of $\langle \sigma_z \rangle$ for the spin-boson model ($s$=0.5, $\omega_c$=10$\Delta$, $\epsilon$=0) in the strong coupling regime for various coupling strength $\alpha$ and temperatures from $T$=0 to $T=\Delta$ is shown. For $T$=0 the dynamics gained from HOPS match very well the dynamics calculated via ML-MCTDH method taken from Ref. \cite{wang_coherent_2010}. The transition from coherent motion to localization upon increasing the coupling strength is nicely seen. Additionally we observe that an increase in temperature dampens the oscillations and causes a relaxation to the non-localized states. Numerical details: A fit of the BCF up to time 15/$\Delta$ with 5 terms yielding an accuracy slightly below 2\% was used. The coupling strength dependent hierarchy depth $k_\mathrm{max}$ was chosen in accordance to Fig. \ref{fig:kmax_conv} as follows: $\alpha$=0.1: 5, $\alpha$=0.15: 6, $\alpha$=0.2: 9, $\alpha$=0.25: 12. To obtain the reduced dynamics the average was taking over $10^5$ stochastic trajectories.}
  \label{fig:thoss_rough}
\end{figure*}

Implementing these considerations allows us obtain converged results for the spin-boson model with sub-Ohmic SD in the strong coupling regime with zero temperature as well as thermal initial bath conditions. For the zero temperature case the dynamics calculated using HOPS (Fig. \ref{fig:thoss_rough} blue line) matches very well the results gained by the ML-MCTDH method (Fig. \ref{fig:thoss_rough} dashed black line) from Ref.\cite{wang_coherent_2010}, reproducing the transition from damped coherent motion at weak coupling to localization upon increasing the coupling strength. Note, the spin dynamics for that particular case has also been calculated using the eHEOM method\cite{duan_zero-temperature_2017}. The influence of a low temperature initial condition ($T$=0.2$\Delta$) is almost negligible for the initial oscillations but becomes evident in the long time behavior suppressing the localization (see insets of Fig. \ref{fig:thoss_rough} for a zoom up on the long time behavior). Further increasing the temperature ($T$=$\Delta$) results also in a stronger damping of the initial oscillations and a faster decay towards the delocalized state. Note that due to the scale-up small fluctuations of the dynamics originating in the stochastic nature of the HOPS method become visible. These fluctuations are unphysical and decrease in amplitude when increasing the number of samples.

\section{Summary and Outlook}

We have shown that the Hierarchy of Pure States (HOPS) method, which provides a general and numerically exact approach to calculate the reduced dynamics for an open quantum system, is very well capable of treating environments with an algebraically decaying bath correlation function corresponding to the class of (sub-) Ohmic spectral densities. HOPS as presented here relies on a representation of the bath correlation function in terms of a sum of exponentials which leads inevitably to an asymptotic exponential decay. However, the algebraic decay can be well approximated over the time interval of interest by minimizing the relative difference between the approximation and the exact bath correlation function directly in the time domain. This is motivated by the NMQSD equation (eq \ref{eqn:nmqsd}) which shows that the reduced dynamics up to time $t$ only depends on values of the bath correlation functions over the interval $[0,t]$. If a good approximative representation for the exact bath correlation function with respect to that particular time interval was found the reduced dynamics obtained from HOPS is exact up to time $t$ irrespectively of the difference in the long time asymptotic behavior. Note that the fitting can be done for any temperature including $T$=0. However, as the fitting procedure might be involved, we have shown that the influence of a thermal initial environmental state can be dealt with by a stochastic Hermitian contribution to the system Hamiltonian for any (not necessarily Hermitian) coupling operator $L$. Additionally we found that in the strong coupling regime this way of incorporating non-zero temperature is even necessary to numerically achieve convergence with respect to the hierarchy depth.

Using the spin-boson model for testing we have compared the dynamics obtained from HOPS with various other methods. In case of week coupling and a fast bath the quantum optical master equation has served as reference. As expected very good agreement was found in the Ohmic case ($s$=1) for various bias values $\epsilon$ and temperatures reaching from $T$=0 to $T$=10$\Delta$. However in the sub-Ohmic case, for non-zero temperature and non-zero bias, deviations occur which are due to the failure of the master equation with time independent rates and which can be cured by introducing time dependent rates\cite{noh_<pre>consistent_2014}.

In the high temperature limit, where the environmental influence can be modeled via a classical random force, we have compared HOPS against two variants of the hierarchical equations of motions (HEOM) method: first the standard variant with a Meier-Tannor decomposition of the spectral density (MT HEOM) and second an extension (eHEOM) capable of treating sub-Ohmic environments correctly\cite{tang_extended_2015}. For parameters where the high temperature limit is applicable, HOPS and eHEOM match very well and coincide with the purely stochastic Hamiltonian approach. The small deviations with respect to HEOM originate in problems with the Meier-Tannor decomposition\cite{tang_extended_2015}.

To treat the strong coupling regime HOPS becomes numerically demanding because the hierarchy depth required for the stochastic state vector to converge increases with the coupling strength. Additionally, the interval for which the exponential representation of the bath correlation function mimics the exact behavior must not be too short resulting in a fair amount of exponential summands. For the HOPS method to converge with respect to the stochastic sampling we have shown again that the non-linear variant is indispensable. Furthermore, when including temperature effects in the strong coupling regime it is highly favorable to treat them by a stochastic Hermitian contribution to the system Hamiltonian part. As HOPS is a stochastic method anyway, this is of no major extra numerical cost. Implementing these considerations allowed us to calculate the dynamics of the spin-boson model with a sub-Ohmic environment ($s$=0.5, $\omega_c$=10$\Delta$) in the strong coupling regime reproducing the ML-MCTDH\cite{wang_coherent_2010} and eHEOM\cite{duan_zero-temperature_2017} results for zero temperature. Further, we successfully used HOPS to calculate the dynamics of a strongly interacting spin with a non-zero temperature environment.

We are convinced that HOPS is widely applicable for open quantum system dynamics. In particular, we have shown that it is suitable to treat environments with (sub-) Ohmic spectral densities. Clearly, as the fitting of the bath correlation function is very generic, it can also be used for many other environments at zero and non-zero temperature. 

Besides investigating particular open quantum systems, HOPS may also serve as input for the transfer tensor method\cite{cerrillo_non-markovian_2014} which allows to efficiently study the long time behavior. In addition to applying HOPS, questions concerning the method itself are left for future work too. As of the very similar structure of HOPS and HEOM details on their relations, in particular for the non-linear HOPS, should be further investigated (see also Süß et al. \cite{suess_hierarchical_2015} where such a relation has already been worked out for the linear HOPS). Hopefully this will shed light on the advantages of each method.

\section{Acknowledgements}

For fruitful discussions about HOPS and anything related to it, we gratefully acknowledge Kimmo Luoma and Alexander Eisfeld. We also thank the International Max Planck Research School at the MPI-PKS Dresden for their support.

The computations were performed on a Bull Cluster at the Center for Information Services and High Performance Computing (ZIH) at TU Dresden. During development and testing of the code frequent use was made of the \texttt{mpmath} library\cite{johansson_<span_2014-1}.

\section*{Appendix}

\appendix
 
\section{Non-zero temperature}

\label{apx_finite_temp}

To incorporate a thermal initial bath state in terms of a stochastic contribution to the system Hamiltonian the reduced density matrix (RDM) at time $t$ is written as
\begin{equation}
  \rho_\sys(t) = \tr_\bath \left[ U(t) \; \rho_\sys(0) \otimes \rho_\beta \; U^\dagger(t) \right]
\end{equation}
where $U(t)$ is the time evolution operator and $\rho_\beta$ the thermal bath state. The P-representation for the thermal bath state reads
\begin{equation}
  \rho_\beta = \bigotimes_\lambda \int \derivD^2 y_\lambda \frac{1}{\pi \bar n_\lambda} e^{-|y_\lambda|^2/ \bar n_\lambda} \proj{y_\lambda}
\end{equation}
with $\bar n_\lambda = (e^{\beta \omega_\lambda} - 1)^{-1}$ being the mean occupation number and $\ket{y_\lambda}$ (normalized) coherent states. These in turn can be written with the help of the displacement operator as $\ket{y_\lambda} = D(y_\lambda)\ket{0}_\lambda$ were $D(y_\lambda) = \exp(y_\lambda a_\lambda^\dagger - y_\lambda^\ast a_\lambda)$. Therefore, the time evolved RDM takes the following form.
\begin{multline}
  \rho_\sys(t) = \bigotimes_\lambda \int \derivD^2 y_\lambda \frac{1}{\pi \bar n_\lambda} e^{-|y_\lambda|^2/ \bar n_\lambda} \tr_\bath \big[ D^\dagger(\ybf)U(t) \; \rho_\sys(0) \\
  \otimes D(\ybf)\proj{\mathbf{0}}D^\dagger(\ybf) \; U^\dagger(t) D(\ybf)\big]
\end{multline}
The boldface is used to indicate the product structure due to the bath modes. Note that the unitarity of the displacement operator $D^\dagger D = 1$ allows for the additional operators under the trace. This expression suggests to introduce a transformed time evolution operator $\tilde U(t) = D^\dagger(\ybf) U(t) D(\ybf)$ with corresponding transformed Hamiltonian
\begin{multline}
  \tilde H = D^\dagger(\ybf) H D(\ybf) = H_\sys + L^\dagger y(t) + L y^\ast(t)\\
  + L^\dagger \sum_\lambda g_\lambda e^{-\im \omega_\lambda t} a_\lambda + \hc
\end{multline}
where $y(t) = \sum_\lambda g_\lambda e^{-\im \omega_\lambda t} y_\lambda$. Note that $H$, being the total Hamiltonian, is already in the interaction picture with respect to the bath. 

We can conclude that the action of the time evolution operator $\tilde U$ mimics the dynamics induced by the Hamiltonian $\tilde H$ which is the original Hamiltonian with the additional Hermitian contribution $L^\dagger y(t) + L y^\ast(t)$ to the system part. Further reading the Gaussian integral in a Monto-Carlo sense yields $\mathbf{E}[y_\lambda] = 0 =\mathbf{E}[y_\lambda y_\lambda']$ and $\mathbf{E}[y_\lambda y^\ast_{\lambda'}] = \bar n_\lambda \delta_{\lambda, \lambda'}$ which in turn gives $\mathbf{E}[y(t)] = 0 = \mathbf{E}[y(t)y(s)]$ and $\mathbf{E}[y(t)y^\ast(s)] = \sum_\lambda \bar n_\lambda |g_\lambda|^2e^{-\im\omega_\lambda(t-s)}$. With that in mind the Gaussian integral can be read as averaging over stochastic processes $y(t)$. Consequently the RDM can be calculated by averaging over stochastic RDMs evolving under the (stochastic) Hamiltonian $\tilde H$ however with initial condition $\rho_\sys(0) \otimes \proj{\mathbf{0}}$. Obviously the evolution of these stochastic RDMs can be obtained using HOPS with zero temperature BCF.

For completeness, given the SD $J(\omega)$ the auto correlation function of the stochastic processes $y(t)$ becomes in the continuous limit:
\begin{equation}
  \mathbf{E}[y(t)y^\ast(s)] = \frac{1}{\pi} \il{0}{\infty}{\omega}\frac{J(\omega)}{e^{\beta\omega}-1} e^{-\im\omega (t-s)}
\end{equation}
Taking the temperature to zero ($\beta \rightarrow \infty$) yields $y(t) = 0$ which eliminates the stochasticity induced by $y(t)$ and reproduced the usual zero temperature expression for the RDM as explained in Sec. \ref{sec_HOPS_derivation}.

\bibliography{exact_dyn_HOPS_arXiv}

\begin{thebibliography}{41}%
\makeatletter
\providecommand \@ifxundefined [1]{%
 \@ifx{#1\undefined}
}%
\providecommand \@ifnum [1]{%
 \ifnum #1\expandafter \@firstoftwo
 \else \expandafter \@secondoftwo
 \fi
}%
\providecommand \@ifx [1]{%
 \ifx #1\expandafter \@firstoftwo
 \else \expandafter \@secondoftwo
 \fi
}%
\providecommand \natexlab [1]{#1}%
\providecommand \enquote  [1]{``#1''}%
\providecommand \bibnamefont  [1]{#1}%
\providecommand \bibfnamefont [1]{#1}%
\providecommand \citenamefont [1]{#1}%
\providecommand \href@noop [0]{\@secondoftwo}%
\providecommand \href [0]{\begingroup \@sanitize@url \@href}%
\providecommand \@href[1]{\@@startlink{#1}\@@href}%
\providecommand \@@href[1]{\endgroup#1\@@endlink}%
\providecommand \@sanitize@url [0]{\catcode `\\12\catcode `\$12\catcode
  `\&12\catcode `\#12\catcode `\^12\catcode `\_12\catcode `\%12\relax}%
\providecommand \@@startlink[1]{}%
\providecommand \@@endlink[0]{}%
\providecommand \url  [0]{\begingroup\@sanitize@url \@url }%
\providecommand \@url [1]{\endgroup\@href {#1}{\urlprefix }}%
\providecommand \urlprefix  [0]{URL }%
\providecommand \Eprint [0]{\href }%
\providecommand \doibase [0]{http://dx.doi.org/}%
\providecommand \selectlanguage [0]{\@gobble}%
\providecommand \bibinfo  [0]{\@secondoftwo}%
\providecommand \bibfield  [0]{\@secondoftwo}%
\providecommand \translation [1]{[#1]}%
\providecommand \BibitemOpen [0]{}%
\providecommand \bibitemStop [0]{}%
\providecommand \bibitemNoStop [0]{.\EOS\space}%
\providecommand \EOS [0]{\spacefactor3000\relax}%
\providecommand \BibitemShut  [1]{\csname bibitem#1\endcsname}%
\let\auto@bib@innerbib\@empty
\bibitem [{\citenamefont {Breuer}\ and\ \citenamefont
  {Petruccione}(2007)}]{breuer_theory_2007}%
  \BibitemOpen
  \bibfield  {author} {\bibinfo {author} {\bibfnamefont {H.-P.}\ \bibnamefont
  {Breuer}}\ and\ \bibinfo {author} {\bibfnamefont {F.}~\bibnamefont
  {Petruccione}},\ }\href@noop {} {\emph {\bibinfo {title} {The {{Theory}} of
  {{Open Quantum Systems}}}}}\ (\bibinfo  {publisher} {{Oxford University
  Press}},\ \bibinfo {address} {Oxford, New York},\ \bibinfo {year}
  {2007})\BibitemShut {NoStop}%
\bibitem [{\citenamefont {Weiss}(2008)}]{weiss_quantum_2008}%
  \BibitemOpen
  \bibfield  {author} {\bibinfo {author} {\bibfnamefont {U.}~\bibnamefont
  {Weiss}},\ }\href@noop {} {\emph {\bibinfo {title} {Quantum {{Dissipative
  Systems}}}}}\ (\bibinfo  {publisher} {{World Scientific}},\ \bibinfo
  {address} {Singapore},\ \bibinfo {year} {2008})\BibitemShut {NoStop}%
\bibitem [{\citenamefont {Rivas}\ and\ \citenamefont
  {Huelga}(2012)}]{rivas_open_2012}%
  \BibitemOpen
  \bibfield  {author} {\bibinfo {author} {\bibfnamefont {A.}~\bibnamefont
  {Rivas}}\ and\ \bibinfo {author} {\bibfnamefont {S.~F.}\ \bibnamefont
  {Huelga}},\ }\href@noop {} {\emph {\bibinfo {title} {Open {{Quantum
  Systems}}. {{An Introduction}}}}}\ (\bibinfo  {publisher} {{Springer}},\
  \bibinfo {address} {Dordrecht London New York},\ \bibinfo {year}
  {2012})\BibitemShut {NoStop}%
\bibitem [{\citenamefont {Strunz}\ \emph {et~al.}(1999)\citenamefont {Strunz},
  \citenamefont {Di{\'o}si},\ and\ \citenamefont {Gisin}}]{strunz_open_1999}%
  \BibitemOpen
  \bibfield  {author} {\bibinfo {author} {\bibfnamefont {W.~T.}\ \bibnamefont
  {Strunz}}, \bibinfo {author} {\bibfnamefont {L.}~\bibnamefont {Di{\'o}si}}, \
  and\ \bibinfo {author} {\bibfnamefont {N.}~\bibnamefont {Gisin}},\ }\href
  {\doibase 10.1103/PhysRevLett.82.1801} {\bibfield  {journal} {\bibinfo
  {journal} {Phys. Rev. Lett.}\ }\textbf {\bibinfo {volume} {82}},\ \bibinfo
  {pages} {1801} (\bibinfo {year} {1999})}\BibitemShut {NoStop}%
\bibitem [{\citenamefont {Suess}\ \emph {et~al.}(2014)\citenamefont {Suess},
  \citenamefont {Eisfeld},\ and\ \citenamefont
  {Strunz}}]{suess_hierarchy_2014}%
  \BibitemOpen
  \bibfield  {author} {\bibinfo {author} {\bibfnamefont {D.}~\bibnamefont
  {Suess}}, \bibinfo {author} {\bibfnamefont {A.}~\bibnamefont {Eisfeld}}, \
  and\ \bibinfo {author} {\bibfnamefont {W.~T.}\ \bibnamefont {Strunz}},\
  }\href {\doibase 10.1103/PhysRevLett.113.150403} {\bibfield  {journal}
  {\bibinfo  {journal} {Phys. Rev. Lett.}\ }\textbf {\bibinfo {volume} {113}},\
  \bibinfo {pages} {150403} (\bibinfo {year} {2014})}\BibitemShut {NoStop}%
\bibitem [{\citenamefont {Tanimura}\ and\ \citenamefont
  {Kubo}(1989)}]{tanimura_two-time_1989}%
  \BibitemOpen
  \bibfield  {author} {\bibinfo {author} {\bibfnamefont {Y.}~\bibnamefont
  {Tanimura}}\ and\ \bibinfo {author} {\bibfnamefont {R.}~\bibnamefont
  {Kubo}},\ }\href {\doibase 10.1143/JPSJ.58.1199} {\bibfield  {journal}
  {\bibinfo  {journal} {J. Phys. Soc. Jpn.}\ }\textbf {\bibinfo {volume}
  {58}},\ \bibinfo {pages} {1199} (\bibinfo {year} {1989})}\BibitemShut
  {NoStop}%
\bibitem [{\citenamefont {Tanimura}(2006)}]{tanimura_stochastic_2006}%
  \BibitemOpen
  \bibfield  {author} {\bibinfo {author} {\bibfnamefont {Y.}~\bibnamefont
  {Tanimura}},\ }\href {\doibase 10.1143/JPSJ.75.082001} {\bibfield  {journal}
  {\bibinfo  {journal} {J. Phys. Soc. Jpn.}\ }\textbf {\bibinfo {volume}
  {75}},\ \bibinfo {pages} {082001} (\bibinfo {year} {2006})}\BibitemShut
  {NoStop}%
\bibitem [{\citenamefont {Makri}\ and\ \citenamefont
  {Makarov}(1995)}]{makri_tensor_1995}%
  \BibitemOpen
  \bibfield  {author} {\bibinfo {author} {\bibfnamefont {N.}~\bibnamefont
  {Makri}}\ and\ \bibinfo {author} {\bibfnamefont {D.~E.}\ \bibnamefont
  {Makarov}},\ }\href {\doibase 10.1063/1.469508} {\bibfield  {journal}
  {\bibinfo  {journal} {J. Chem. Phys.}\ }\textbf {\bibinfo {volume} {102}},\
  \bibinfo {pages} {4600} (\bibinfo {year} {1995})}\BibitemShut {NoStop}%
\bibitem [{\citenamefont {Orth}\ \emph {et~al.}(2013)\citenamefont {Orth},
  \citenamefont {Imambekov},\ and\ \citenamefont
  {Le~Hur}}]{orth_nonperturbative_2013}%
  \BibitemOpen
  \bibfield  {author} {\bibinfo {author} {\bibfnamefont {P.~P.}\ \bibnamefont
  {Orth}}, \bibinfo {author} {\bibfnamefont {A.}~\bibnamefont {Imambekov}}, \
  and\ \bibinfo {author} {\bibfnamefont {K.}~\bibnamefont {Le~Hur}},\ }\href
  {\doibase 10.1103/PhysRevB.87.014305} {\bibfield  {journal} {\bibinfo
  {journal} {Phys. Rev. B}\ }\textbf {\bibinfo {volume} {87}},\ \bibinfo
  {pages} {014305} (\bibinfo {year} {2013})}\BibitemShut {NoStop}%
\bibitem [{\citenamefont {Meyer}\ \emph {et~al.}(1990)\citenamefont {Meyer},
  \citenamefont {Manthe},\ and\ \citenamefont
  {Cederbaum}}]{meyer_multi-configurational_1990}%
  \BibitemOpen
  \bibfield  {author} {\bibinfo {author} {\bibfnamefont {H.~D.}\ \bibnamefont
  {Meyer}}, \bibinfo {author} {\bibfnamefont {U.}~\bibnamefont {Manthe}}, \
  and\ \bibinfo {author} {\bibfnamefont {L.~S.}\ \bibnamefont {Cederbaum}},\
  }\href {\doibase 10.1016/0009-2614(90)87014-I} {\bibfield  {journal}
  {\bibinfo  {journal} {Chem. Phys. Lett.}\ }\textbf {\bibinfo {volume}
  {165}},\ \bibinfo {pages} {73} (\bibinfo {year} {1990})}\BibitemShut
  {NoStop}%
\bibitem [{\citenamefont {Beck}\ \emph {et~al.}(2000)\citenamefont {Beck},
  \citenamefont {J{\"a}ckle}, \citenamefont {Worth},\ and\ \citenamefont
  {Meyer}}]{beck_multiconfiguration_2000}%
  \BibitemOpen
  \bibfield  {author} {\bibinfo {author} {\bibfnamefont {M.~H.}\ \bibnamefont
  {Beck}}, \bibinfo {author} {\bibfnamefont {A.}~\bibnamefont {J{\"a}ckle}},
  \bibinfo {author} {\bibfnamefont {G.~A.}\ \bibnamefont {Worth}}, \ and\
  \bibinfo {author} {\bibfnamefont {H.~D.}\ \bibnamefont {Meyer}},\ }\href
  {\doibase 10.1016/S0370-1573(99)00047-2} {\bibfield  {journal} {\bibinfo
  {journal} {Phys. Rep.}\ }\textbf {\bibinfo {volume} {324}},\ \bibinfo {pages}
  {1} (\bibinfo {year} {2000})}\BibitemShut {NoStop}%
\bibitem [{\citenamefont {Wang}\ and\ \citenamefont
  {Thoss}(2003)}]{wang_multilayer_2003}%
  \BibitemOpen
  \bibfield  {author} {\bibinfo {author} {\bibfnamefont {H.}~\bibnamefont
  {Wang}}\ and\ \bibinfo {author} {\bibfnamefont {M.}~\bibnamefont {Thoss}},\
  }\href {\doibase 10.1063/1.1580111} {\bibfield  {journal} {\bibinfo
  {journal} {J. Chem. Phys.}\ }\textbf {\bibinfo {volume} {119}},\ \bibinfo
  {pages} {1289} (\bibinfo {year} {2003})}\BibitemShut {NoStop}%
\bibitem [{\citenamefont {Thorwart}\ \emph {et~al.}(2004)\citenamefont
  {Thorwart}, \citenamefont {Paladino},\ and\ \citenamefont
  {Grifoni}}]{thorwart_dynamics_2004}%
  \BibitemOpen
  \bibfield  {author} {\bibinfo {author} {\bibfnamefont {M.}~\bibnamefont
  {Thorwart}}, \bibinfo {author} {\bibfnamefont {E.}~\bibnamefont {Paladino}},
  \ and\ \bibinfo {author} {\bibfnamefont {M.}~\bibnamefont {Grifoni}},\ }\href
  {\doibase 10.1016/j.chemphys.2003.10.007} {\bibfield  {journal} {\bibinfo
  {journal} {Chem. Phys.}\ }\textbf {\bibinfo {volume} {296}},\ \bibinfo
  {pages} {333} (\bibinfo {year} {2004})}\BibitemShut {NoStop}%
\bibitem [{\citenamefont {Nalbach}\ and\ \citenamefont
  {Thorwart}(2010)}]{nalbach_ultraslow_2010}%
  \BibitemOpen
  \bibfield  {author} {\bibinfo {author} {\bibfnamefont {P.}~\bibnamefont
  {Nalbach}}\ and\ \bibinfo {author} {\bibfnamefont {M.}~\bibnamefont
  {Thorwart}},\ }\href {\doibase 10.1103/PhysRevB.81.054308} {\bibfield
  {journal} {\bibinfo  {journal} {Phys. Rev. B}\ }\textbf {\bibinfo {volume}
  {81}},\ \bibinfo {pages} {054308} (\bibinfo {year} {2010})}\BibitemShut
  {NoStop}%
\bibitem [{\citenamefont {Tanimura}(2014)}]{tanimura_reduced_2014}%
  \BibitemOpen
  \bibfield  {author} {\bibinfo {author} {\bibfnamefont {Y.}~\bibnamefont
  {Tanimura}},\ }\href {\doibase 10.1063/1.4890441} {\bibfield  {journal}
  {\bibinfo  {journal} {J. Chem. Phys.}\ }\textbf {\bibinfo {volume} {141}},\
  \bibinfo {pages} {044114} (\bibinfo {year} {2014})}\BibitemShut {NoStop}%
\bibitem [{\citenamefont {Wang}\ and\ \citenamefont
  {Thoss}(2008)}]{wang_coherent_2008}%
  \BibitemOpen
  \bibfield  {author} {\bibinfo {author} {\bibfnamefont {H.}~\bibnamefont
  {Wang}}\ and\ \bibinfo {author} {\bibfnamefont {M.}~\bibnamefont {Thoss}},\
  }\href {\doibase 10.1088/1367-2630/10/11/115005} {\bibfield  {journal}
  {\bibinfo  {journal} {New J. Phys.}\ }\textbf {\bibinfo {volume} {10}},\
  \bibinfo {pages} {115005} (\bibinfo {year} {2008})}\BibitemShut {NoStop}%
\bibitem [{\citenamefont {Wang}\ and\ \citenamefont
  {Thoss}(2010)}]{wang_coherent_2010}%
  \BibitemOpen
  \bibfield  {author} {\bibinfo {author} {\bibfnamefont {H.}~\bibnamefont
  {Wang}}\ and\ \bibinfo {author} {\bibfnamefont {M.}~\bibnamefont {Thoss}},\
  }\href {\doibase 10.1016/j.chemphys.2010.02.027} {\bibfield  {journal}
  {\bibinfo  {journal} {Chem. Phys.}\ }\bibinfo {series} {Dynamics of molecular
  systems: From quantum to classical},\ \textbf {\bibinfo {volume} {370}},\
  \bibinfo {pages} {78} (\bibinfo {year} {2010})}\BibitemShut {NoStop}%
\bibitem [{\citenamefont {Matzkies}\ and\ \citenamefont
  {Manthe}(1998)}]{matzkies_accurate_1998}%
  \BibitemOpen
  \bibfield  {author} {\bibinfo {author} {\bibfnamefont {F.}~\bibnamefont
  {Matzkies}}\ and\ \bibinfo {author} {\bibfnamefont {U.}~\bibnamefont
  {Manthe}},\ }\href {\doibase 10.1063/1.478128} {\bibfield  {journal}
  {\bibinfo  {journal} {J. Chem. Phys.}\ }\textbf {\bibinfo {volume} {110}},\
  \bibinfo {pages} {88} (\bibinfo {year} {1998})}\BibitemShut {NoStop}%
\bibitem [{\citenamefont {Wang}\ and\ \citenamefont
  {Thoss}(2006)}]{wang_quantum-mechanical_2006}%
  \BibitemOpen
  \bibfield  {author} {\bibinfo {author} {\bibfnamefont {H.}~\bibnamefont
  {Wang}}\ and\ \bibinfo {author} {\bibfnamefont {M.}~\bibnamefont {Thoss}},\
  }\href {\doibase 10.1063/1.2161178} {\bibfield  {journal} {\bibinfo
  {journal} {J. Chem. Phys.}\ }\textbf {\bibinfo {volume} {124}},\ \bibinfo
  {pages} {034114} (\bibinfo {year} {2006})}\BibitemShut {NoStop}%
\bibitem [{\citenamefont {Meier}\ and\ \citenamefont
  {Tannor}(1999)}]{meier_non-markovian_1999}%
  \BibitemOpen
  \bibfield  {author} {\bibinfo {author} {\bibfnamefont {C.}~\bibnamefont
  {Meier}}\ and\ \bibinfo {author} {\bibfnamefont {D.~J.}\ \bibnamefont
  {Tannor}},\ }\href {\doibase 10.1063/1.479669} {\bibfield  {journal}
  {\bibinfo  {journal} {J. Chem. Phys.}\ }\textbf {\bibinfo {volume} {111}},\
  \bibinfo {pages} {3365} (\bibinfo {year} {1999})}\BibitemShut {NoStop}%
\bibitem [{\citenamefont {Liu}\ \emph {et~al.}(2014)\citenamefont {Liu},
  \citenamefont {Zhu}, \citenamefont {Bai},\ and\ \citenamefont
  {Shi}}]{liu_reduced_2014}%
  \BibitemOpen
  \bibfield  {author} {\bibinfo {author} {\bibfnamefont {H.}~\bibnamefont
  {Liu}}, \bibinfo {author} {\bibfnamefont {L.}~\bibnamefont {Zhu}}, \bibinfo
  {author} {\bibfnamefont {S.}~\bibnamefont {Bai}}, \ and\ \bibinfo {author}
  {\bibfnamefont {Q.}~\bibnamefont {Shi}},\ }\href {\doibase 10.1063/1.4870035}
  {\bibfield  {journal} {\bibinfo  {journal} {J. Chem. Phys.}\ }\textbf
  {\bibinfo {volume} {140}},\ \bibinfo {pages} {134106} (\bibinfo {year}
  {2014})}\BibitemShut {NoStop}%
\bibitem [{\citenamefont {Tang}\ \emph {et~al.}(2015)\citenamefont {Tang},
  \citenamefont {Ouyang}, \citenamefont {Gong}, \citenamefont {Wang},\ and\
  \citenamefont {Wu}}]{tang_extended_2015}%
  \BibitemOpen
  \bibfield  {author} {\bibinfo {author} {\bibfnamefont {Z.}~\bibnamefont
  {Tang}}, \bibinfo {author} {\bibfnamefont {X.}~\bibnamefont {Ouyang}},
  \bibinfo {author} {\bibfnamefont {Z.}~\bibnamefont {Gong}}, \bibinfo {author}
  {\bibfnamefont {H.}~\bibnamefont {Wang}}, \ and\ \bibinfo {author}
  {\bibfnamefont {J.}~\bibnamefont {Wu}},\ }\href {\doibase 10.1063/1.4936924}
  {\bibfield  {journal} {\bibinfo  {journal} {J. Chem. Phys.}\ }\textbf
  {\bibinfo {volume} {143}},\ \bibinfo {pages} {224112} (\bibinfo {year}
  {2015})}\BibitemShut {NoStop}%
\bibitem [{\citenamefont {Li}\ \emph {et~al.}(2012)\citenamefont {Li},
  \citenamefont {Tong}, \citenamefont {Zheng}, \citenamefont {Hou},
  \citenamefont {Wei}, \citenamefont {Hu},\ and\ \citenamefont
  {Yan}}]{li_hierarchical_2012}%
  \BibitemOpen
  \bibfield  {author} {\bibinfo {author} {\bibfnamefont {Z.}~\bibnamefont
  {Li}}, \bibinfo {author} {\bibfnamefont {N.}~\bibnamefont {Tong}}, \bibinfo
  {author} {\bibfnamefont {X.}~\bibnamefont {Zheng}}, \bibinfo {author}
  {\bibfnamefont {D.}~\bibnamefont {Hou}}, \bibinfo {author} {\bibfnamefont
  {J.}~\bibnamefont {Wei}}, \bibinfo {author} {\bibfnamefont {J.}~\bibnamefont
  {Hu}}, \ and\ \bibinfo {author} {\bibfnamefont {Y.}~\bibnamefont {Yan}},\
  }\href {\doibase 10.1103/PhysRevLett.109.266403} {\bibfield  {journal}
  {\bibinfo  {journal} {Phys. Rev. Lett.}\ }\textbf {\bibinfo {volume} {109}},\
  \bibinfo {pages} {266403} (\bibinfo {year} {2012})}\BibitemShut {NoStop}%
\bibitem [{\citenamefont {Cheng}\ \emph {et~al.}(2015)\citenamefont {Cheng},
  \citenamefont {Hou}, \citenamefont {Wang}, \citenamefont {Li}, \citenamefont
  {Wei},\ and\ \citenamefont {Yan}}]{cheng_time-dependent_2015}%
  \BibitemOpen
  \bibfield  {author} {\bibinfo {author} {\bibfnamefont {Y.}~\bibnamefont
  {Cheng}}, \bibinfo {author} {\bibfnamefont {W.}~\bibnamefont {Hou}}, \bibinfo
  {author} {\bibfnamefont {Y.}~\bibnamefont {Wang}}, \bibinfo {author}
  {\bibfnamefont {Z.}~\bibnamefont {Li}}, \bibinfo {author} {\bibfnamefont
  {J.}~\bibnamefont {Wei}}, \ and\ \bibinfo {author} {\bibfnamefont
  {Y.}~\bibnamefont {Yan}},\ }\href {\doibase 10.1088/1367-2630/17/3/033009}
  {\bibfield  {journal} {\bibinfo  {journal} {New J. Phys.}\ }\textbf {\bibinfo
  {volume} {17}},\ \bibinfo {pages} {033009} (\bibinfo {year}
  {2015})}\BibitemShut {NoStop}%
\bibitem [{\citenamefont {Duan}\ \emph {et~al.}(2017)\citenamefont {Duan},
  \citenamefont {Tang}, \citenamefont {Cao},\ and\ \citenamefont
  {Wu}}]{duan_zero-temperature_2017}%
  \BibitemOpen
  \bibfield  {author} {\bibinfo {author} {\bibfnamefont {C.}~\bibnamefont
  {Duan}}, \bibinfo {author} {\bibfnamefont {Z.}~\bibnamefont {Tang}}, \bibinfo
  {author} {\bibfnamefont {J.}~\bibnamefont {Cao}}, \ and\ \bibinfo {author}
  {\bibfnamefont {J.}~\bibnamefont {Wu}},\ }\href {\doibase
  10.1103/PhysRevB.95.214308} {\bibfield  {journal} {\bibinfo  {journal} {Phys.
  Rev. B}\ }\textbf {\bibinfo {volume} {95}},\ \bibinfo {pages} {214308}
  (\bibinfo {year} {2017})}\BibitemShut {NoStop}%
\bibitem [{\citenamefont {Zhang}\ and\ \citenamefont
  {Eisfeld}(2016)}]{zhang_non-perturbative_2016}%
  \BibitemOpen
  \bibfield  {author} {\bibinfo {author} {\bibfnamefont {P.-P.}\ \bibnamefont
  {Zhang}}\ and\ \bibinfo {author} {\bibfnamefont {A.}~\bibnamefont
  {Eisfeld}},\ }\href {\doibase 10.1021/acs.jpclett.6b02111} {\bibfield
  {journal} {\bibinfo  {journal} {J. Phys. Chem. Lett.}\ }\textbf {\bibinfo
  {volume} {7}},\ \bibinfo {pages} {4488} (\bibinfo {year} {2016})}\BibitemShut
  {NoStop}%
\bibitem [{\citenamefont {Di{\'o}si}\ and\ \citenamefont
  {Strunz}(1997)}]{diosi_non-markovian_1997}%
  \BibitemOpen
  \bibfield  {author} {\bibinfo {author} {\bibfnamefont {L.}~\bibnamefont
  {Di{\'o}si}}\ and\ \bibinfo {author} {\bibfnamefont {W.~T.}\ \bibnamefont
  {Strunz}},\ }\href {\doibase 10.1016/S0375-9601(97)00717-2} {\bibfield
  {journal} {\bibinfo  {journal} {Phys. Lett. A}\ }\textbf {\bibinfo {volume}
  {235}},\ \bibinfo {pages} {569} (\bibinfo {year} {1997})}\BibitemShut
  {NoStop}%
\bibitem [{\citenamefont {Noh}\ and\ \citenamefont
  {Fischer}(2014)}]{noh_<pre>consistent_2014}%
  \BibitemOpen
  \bibfield  {author} {\bibinfo {author} {\bibfnamefont {K.-J.}\ \bibnamefont
  {Noh}}\ and\ \bibinfo {author} {\bibfnamefont {U.~R.}\ \bibnamefont
  {Fischer}},\ }\href {\doibase 10.1103/PhysRevB.90.220302} {\bibfield
  {journal} {\bibinfo  {journal} {Phys. Rev. B}\ }\textbf {\bibinfo {volume}
  {90}},\ \bibinfo {pages} {220302} (\bibinfo {year} {2014})}\BibitemShut
  {NoStop}%
\bibitem [{\citenamefont {{de Vega}}\ \emph {et~al.}(2005)\citenamefont {{de
  Vega}}, \citenamefont {Alonso}, \citenamefont {Gaspard},\ and\ \citenamefont
  {Strunz}}]{de_vega_non-markovian_2005}%
  \BibitemOpen
  \bibfield  {author} {\bibinfo {author} {\bibfnamefont {I.}~\bibnamefont {{de
  Vega}}}, \bibinfo {author} {\bibfnamefont {D.}~\bibnamefont {Alonso}},
  \bibinfo {author} {\bibfnamefont {P.}~\bibnamefont {Gaspard}}, \ and\
  \bibinfo {author} {\bibfnamefont {W.~T.}\ \bibnamefont {Strunz}},\ }\href
  {\doibase 10.1063/1.1867377} {\bibfield  {journal} {\bibinfo  {journal} {J.
  Chem. Phys.}\ }\textbf {\bibinfo {volume} {122}},\ \bibinfo {pages} {124106}
  (\bibinfo {year} {2005})}\BibitemShut {NoStop}%
\bibitem [{\citenamefont {Moix}\ and\ \citenamefont
  {Cao}(2013)}]{moix_hybrid_2013}%
  \BibitemOpen
  \bibfield  {author} {\bibinfo {author} {\bibfnamefont {J.~M.}\ \bibnamefont
  {Moix}}\ and\ \bibinfo {author} {\bibfnamefont {J.}~\bibnamefont {Cao}},\
  }\href {\doibase 10.1063/1.4822043} {\bibfield  {journal} {\bibinfo
  {journal} {J. Chem. Phys.}\ }\textbf {\bibinfo {volume} {139}},\ \bibinfo
  {pages} {134106} (\bibinfo {year} {2013})}\BibitemShut {NoStop}%
\bibitem [{\citenamefont {Ritschel}\ and\ \citenamefont
  {Eisfeld}(2014)}]{ritschel_analytic_2014}%
  \BibitemOpen
  \bibfield  {author} {\bibinfo {author} {\bibfnamefont {G.}~\bibnamefont
  {Ritschel}}\ and\ \bibinfo {author} {\bibfnamefont {A.}~\bibnamefont
  {Eisfeld}},\ }\href {\doibase 10.1063/1.4893931} {\bibfield  {journal}
  {\bibinfo  {journal} {J. Chem. Phys.}\ }\textbf {\bibinfo {volume} {141}},\
  \bibinfo {pages} {094101} (\bibinfo {year} {2014})}\BibitemShut {NoStop}%
\bibitem [{\citenamefont {Stockburger}\ and\ \citenamefont
  {Grabert}(2002)}]{stockburger_<pre>exact_2002}%
  \BibitemOpen
  \bibfield  {author} {\bibinfo {author} {\bibfnamefont {J.~T.}\ \bibnamefont
  {Stockburger}}\ and\ \bibinfo {author} {\bibfnamefont {H.}~\bibnamefont
  {Grabert}},\ }\href {\doibase 10.1103/PhysRevLett.88.170407} {\bibfield
  {journal} {\bibinfo  {journal} {Phys. Rev. Lett.}\ }\textbf {\bibinfo
  {volume} {88}},\ \bibinfo {pages} {170407} (\bibinfo {year}
  {2002})}\BibitemShut {NoStop}%
\bibitem [{\citenamefont {Chen}\ \emph {et~al.}(2013)\citenamefont {Chen},
  \citenamefont {Cao},\ and\ \citenamefont {Silbey}}]{chen_novel_2013}%
  \BibitemOpen
  \bibfield  {author} {\bibinfo {author} {\bibfnamefont {X.}~\bibnamefont
  {Chen}}, \bibinfo {author} {\bibfnamefont {J.}~\bibnamefont {Cao}}, \ and\
  \bibinfo {author} {\bibfnamefont {R.~J.}\ \bibnamefont {Silbey}},\ }\href
  {\doibase 10.1063/1.4808377} {\bibfield  {journal} {\bibinfo  {journal} {J.
  Chem. Phys.}\ }\textbf {\bibinfo {volume} {138}},\ \bibinfo {pages} {224104}
  (\bibinfo {year} {2013})}\BibitemShut {NoStop}%
\bibitem [{\citenamefont {Jones}\ \emph {et~al.}(2017)\citenamefont {Jones},
  \citenamefont {Oliphant}, \citenamefont {Peterson},\ and\ \citenamefont
  {{others}}}]{jones_scipy:_2017}%
  \BibitemOpen
  \bibfield  {author} {\bibinfo {author} {\bibfnamefont {E.}~\bibnamefont
  {Jones}}, \bibinfo {author} {\bibfnamefont {T.}~\bibnamefont {Oliphant}},
  \bibinfo {author} {\bibfnamefont {P.}~\bibnamefont {Peterson}}, \ and\
  \bibinfo {author} {\bibnamefont {{others}}},\ }\href@noop {} {\emph {\bibinfo
  {title} {{{SciPy}}: {{Open}} Source Scientific Tools for {{Python}}}}}\
  (\bibinfo  {publisher} {{\url{www.scipy.org}}},\ \bibinfo {year}
  {2017})\BibitemShut {NoStop}%
\bibitem [{\citenamefont {Hartmann}(2016)}]{hartmann_stocproc:_2016}%
  \BibitemOpen
  \bibfield  {author} {\bibinfo {author} {\bibfnamefont {R.}~\bibnamefont
  {Hartmann}},\ }\href@noop {} {\emph {\bibinfo {title} {{{StocProc}}: {{Time}}
  Continuous Stochastic Process Generator for {{Python}}}}}\ (\bibinfo
  {publisher} {{\url{www.github.com/cimatosa/stocproc}}},\ \bibinfo {year}
  {2016})\BibitemShut {NoStop}%
\bibitem [{\citenamefont {Leggett}\ \emph {et~al.}(1987)\citenamefont
  {Leggett}, \citenamefont {Chakravarty}, \citenamefont {Dorsey}, \citenamefont
  {Fisher}, \citenamefont {Garg},\ and\ \citenamefont
  {Zwerger}}]{leggett_dynamics_1987}%
  \BibitemOpen
  \bibfield  {author} {\bibinfo {author} {\bibfnamefont {A.~J.}\ \bibnamefont
  {Leggett}}, \bibinfo {author} {\bibfnamefont {S.}~\bibnamefont
  {Chakravarty}}, \bibinfo {author} {\bibfnamefont {A.~T.}\ \bibnamefont
  {Dorsey}}, \bibinfo {author} {\bibfnamefont {M.~P.~A.}\ \bibnamefont
  {Fisher}}, \bibinfo {author} {\bibfnamefont {A.}~\bibnamefont {Garg}}, \ and\
  \bibinfo {author} {\bibfnamefont {W.}~\bibnamefont {Zwerger}},\ }\href
  {\doibase 10.1103/RevModPhys.59.1} {\bibfield  {journal} {\bibinfo  {journal}
  {Rev. Mod. Phys.}\ }\textbf {\bibinfo {volume} {59}},\ \bibinfo {pages} {1}
  (\bibinfo {year} {1987})}\BibitemShut {NoStop}%
\bibitem [{\citenamefont {Kehrein}\ and\ \citenamefont
  {Mielke}(1996)}]{kehrein_spin-boson_1996}%
  \BibitemOpen
  \bibfield  {author} {\bibinfo {author} {\bibfnamefont {S.~K.}\ \bibnamefont
  {Kehrein}}\ and\ \bibinfo {author} {\bibfnamefont {A.}~\bibnamefont
  {Mielke}},\ }\href {\doibase 10.1016/0375-9601(96)00475-6} {\bibfield
  {journal} {\bibinfo  {journal} {Phys. Lett. A}\ }\textbf {\bibinfo {volume}
  {219}},\ \bibinfo {pages} {313} (\bibinfo {year} {1996})}\BibitemShut
  {NoStop}%
\bibitem [{\citenamefont {Shnirman}\ and\ \citenamefont
  {Sch{\"o}n}(2003)}]{shnirman_dephasing_2003}%
  \BibitemOpen
  \bibfield  {author} {\bibinfo {author} {\bibfnamefont {A.}~\bibnamefont
  {Shnirman}}\ and\ \bibinfo {author} {\bibfnamefont {G.}~\bibnamefont
  {Sch{\"o}n}},\ }in\ \href {\doibase 10.1007/978-94-010-0089-5_17} {\emph
  {\bibinfo {booktitle} {Quantum {{Noise}} in {{Mesoscopic Physics}}}}},\
  \bibinfo {series and number} {\bibinfo {series} {NATO Science Series}\
  No.~\bibinfo {number} {97}},\ \bibinfo {editor} {edited by\ \bibinfo {editor}
  {\bibfnamefont {Y.~V.}\ \bibnamefont {Nazarov}}}\ (\bibinfo  {publisher}
  {{Springer}},\ \bibinfo {address} {Netherlands},\ \bibinfo {year} {2003})\
  pp.\ \bibinfo {pages} {357--370}\BibitemShut {NoStop}%
\bibitem [{\citenamefont {Cerrillo}\ and\ \citenamefont
  {Cao}(2014)}]{cerrillo_non-markovian_2014}%
  \BibitemOpen
  \bibfield  {author} {\bibinfo {author} {\bibfnamefont {J.}~\bibnamefont
  {Cerrillo}}\ and\ \bibinfo {author} {\bibfnamefont {J.}~\bibnamefont {Cao}},\
  }\href {\doibase 10.1103/PhysRevLett.112.110401} {\bibfield  {journal}
  {\bibinfo  {journal} {Phys. Rev. Lett.}\ }\textbf {\bibinfo {volume} {112}},\
  \bibinfo {pages} {110401} (\bibinfo {year} {2014})}\BibitemShut {NoStop}%
\bibitem [{\citenamefont {Suess}\ \emph {et~al.}(2015)\citenamefont {Suess},
  \citenamefont {Strunz},\ and\ \citenamefont
  {Eisfeld}}]{suess_hierarchical_2015}%
  \BibitemOpen
  \bibfield  {author} {\bibinfo {author} {\bibfnamefont {D.}~\bibnamefont
  {Suess}}, \bibinfo {author} {\bibfnamefont {W.~T.}\ \bibnamefont {Strunz}}, \
  and\ \bibinfo {author} {\bibfnamefont {A.}~\bibnamefont {Eisfeld}},\ }\href
  {\doibase 10.1007/s10955-015-1236-7} {\bibfield  {journal} {\bibinfo
  {journal} {J. Stat. Phys.}\ }\textbf {\bibinfo {volume} {159}},\ \bibinfo
  {pages} {1408} (\bibinfo {year} {2015})}\BibitemShut {NoStop}%
\bibitem [{\citenamefont {Johansson}\ and\ \citenamefont
  {{others}}(2014)}]{johansson_<span_2014-1}%
  \BibitemOpen
  \bibfield  {author} {\bibinfo {author} {\bibfnamefont {F.}~\bibnamefont
  {Johansson}}\ and\ \bibinfo {author} {\bibnamefont {{others}}},\ }\href@noop
  {} {\emph {\bibinfo {title} {{{mpmath}}: A {{Python}} Library for
  Arbitrary-Precision Floating-Point Arithmetic}}}\ (\bibinfo  {publisher}
  {{\url{www.mpmath.org}}},\ \bibinfo {year} {2014})\BibitemShut {NoStop}%
\end{thebibliography}%

\end{document}